\documentclass[a4paper,12pt]{article}
\usepackage{pos}

\usepackage{dsfont,hyperref}  %%FRK to get clickable refs use PDFLaTeX

\usepackage{graphicx} % Required for inserting images
\usepackage{epstopdf} % Required if using pdfLaTeX to automatically convert EPS to PDF

\newcommand {\version}{v4}
\newcommand{\beq}{\begin{equation}}
\newcommand{\eeq}{\end{equation}}
\newcommand{\beqa}{\begin{eqnarray}}
\newcommand{\eeqa}{\end{eqnarray}}
\newcommand{\beqNO}{\begin{equation*}}
\newcommand{\eeqNO}{\end{equation*}}
\newcommand{\beqaNO}{\begin{eqnarray*}}
\newcommand{\eeqaNO}{\end{eqnarray*}}
\newcommand{\bsubeqs}{\begin{subequations}}
\newcommand{\esubeqs}{\end{subequations}}

\title{Big Bang revisited}
\author*[\,a]{Frans R. Klinkhamer}  %%FRK: extra space before suffix

\affiliation[a]{Institute for Theoretical Physics,
Karlsruhe Institute of Technology (KIT),
76128 Karlsruhe, Germany}

\emailAdd{frans.klinkhamer@kit.edu; frans.klinkhamer@gmail.com}

  \abstract{The Friedmann cosmological solution
of the standard Einstein gravitational
field equation has a curvature singularity
at a moment in time known as the Big Bang.
It has been suggested that
this Big Bang curvature singularity
can be eliminated
by use of a degenerate spacetime metric.
This proposal was the main topic of
our talk at the Workshop,
but, here, we also discuss
the possible appearance
of CPT-conjugated worlds and the conjectured
relevance of an extended version of
Einstein's field equation.
\newline\\  \\arXiv:2604.00077 \hfill preprint KA--TP--08--2026 (\version) %%FRK
}

\FullConference{Proceedings of the Corfu Summer Institute
2025 "School and Workshops on Elementary Particle Physics
and Gravity" (CORFU2025)\\
27 April - 28 September, 2025\\
Corfu, Greece\\}

\begin{document}
\maketitle

\section{Introduction}
\label{sec:Introduction}

It is important to realize, right from the start (!), that
the so-called Big Bang is a \emph{theoretical concept},
not the result of direct astronomical observations.
In fact, the Big Bang  (BB)
follows from an extrapolation back
in cosmic time.
This ``journey-back-in-time'' was
initiated by Friedmann, later joined by Lema\^{i}tre,
Gamov, and others.

The well-known Friedmann solution~\cite{Friedmann1922-24},
with a present moment ($t=t_{0}$) of
expansion (i.e., decreasing energy density)
as discovered by Hubble~\cite{Hubble1929},
extrapolates back in time to a moment
($t=t_\text{BB} \equiv 0 < t_{0}$)
when the Universe had a diverging spacetime
curvature (and also an infinite energy density).
That moment is called the Big Bang.

An alternative cosmological solution
without curvature singularity has been
proposed recently~\cite{Klinkhamer2019}
and was the topic of our talk
at the 2025 Corfu Workshop on Tensions in Cosmology.
As there has already appeared an
extensive review of this alternative solution
in Ref.~\cite{Klinkhamer2025-MPLAreview},
the present contribution to the Proceedings can be relatively
short. But we will add two topics %%sections
not covered in the previous review,
the first (Sec.~\ref{sec:Defect-cosmology-FLC-universe})
addresses the
possibility of having CPT-conjugated
twin worlds and the second
(Sec.~\ref{sec:Behavior-at-tis0})
discusses the behavior at the defect hypersurface
and
the possible relevance of an
``extended'' Einstein field equation,
already suggested by Einstein and Rosen
in a different context~\cite{EinsteinRosen1935}.

Throughout, we use the metric
signature $(-+++)$ and,
unless stated otherwise, natural units
with $c=1$ and $\hbar=1$.

%%\newpage%%tmp
\section{Friedmann cosmology and a question}
\label{sec:Friedmann-cosmology}

A particular Friedmann
solution~\cite{Friedmann1922-24}
for the cosmic scale factor $a(t)$ and
an ultra-relativistic
energy density $\rho_{M}(t)$ reads:%
\bsubeqs\label{eq:Friedmann-solution}
\beqa
\label{eq:Friedmann-asol}
a(t)\,\Big|_\text{F}^{(w_{M}=1/3)}
&=&
\sqrt{\left(t-t_\text{BB}\right)/\left(t_{0}-t_\text{BB}\right)}\,,
\hspace*{8mm}
\text{for}\;\;\;t,\,t_{0}>t_\text{BB} \equiv 0\,,
\\[1mm]
\label{eq:Friedmann-rhoMsol}
\rho_{M}(t)\,\Big|_\text{F}^{(w_{M}=1/3)}
&=&
3\,P_{M}(t)\,\Big|_\text{F}^{(w_{M}=1/3)}
%%=\rho_{M,0}/a^{4}(t)\,\Big|_\text{F}^{(w_{M}=1/3)}
=\rho_{M,0}\,
\left(t_{0}-t_\text{BB}\right)^{2}/
\left(t-t_\text{BB}\right)^{2}\,,
\eeqa
\esubeqs
with boundary condition $H_{0}$ $\equiv$
$\big[\big(\mathrm{d} a(t)/\mathrm{d} t\big)\big/a(t)\,\big]_{t=t_{0}}> 0$
for $t_{0} > t_\text{BB} \equiv 0$,
according to Hubble~\cite{Hubble1929}.
Here, time translation invariance
has been used to set $t_\text{BB}$ to zero
in the solution.

The solution \eqref{eq:Friedmann-solution}
is obtained
from the Einstein gravitational field
equation (cf. the
textbooks \cite{Weinberg1972,HawkingEllis1973})
\begin{equation}
\label{eq:Einstein-eq}
R_{\mu\nu}= 8\pi G\;
\left(T_{\mu\nu}^\text{\,(M)}
- \frac12\, g_{\mu\nu}\,T^\text{\,(M)}\right)
\end{equation}
by inserting an appropriate metric \textit{Ansatz} and
making a suitable choice for the matter content.
Specifically, we take
the homogeneous and isotropic
spatially-flat Robertson--Walker metric (cf.
Refs.~\cite{Weinberg1972,HawkingEllis1973})
\bsubeqs\label{eq:RW}
\beqa\label{eq:RW-ds2}
\hspace*{0mm}
ds^{2}\,\Big|_\text{RW}
&\equiv&
g_{\mu\nu}(x)\, dx^{\mu}\,dx^{\nu} \,
\Big|_\text{RW}
%\nonumber\\[1mm]
=
- d t^{2}
+ a^{2}( t )\;
\Big[ \big(d x^{1}\big)^{2} +
\big(d x^{2}\big)^{2} +\big(d x^{3}\big)^{2} \Big]\,,
\\[1mm]
x^{1},\,x^{2},\,x^{3} &\in& (-\infty,\,\infty)
\,,
\eeqa
\esubeqs
for Cartesian spatial coordinates
$\{x^{1},\,x^{2},\,x^{3}\} \in \mathbb{R}^{3}$
and cosmic time coordinate $x^{0}=t \in \mathbb{R}$
(with a range restricted by the dynamics, as will
become clear shortly).
For the matter content, we take the energy-momentum
tensor $T_{\mu\nu}^\text{\,(M)}$ of a
homogeneous relativistic perfect fluid, with
energy density $\rho_{M}(t)$
and pressure $P_{M}(t) =\rho_{M}(t)/3$.
The simple case considered has, thus,
a \emph{constant} equation-of-state %%(EOS)
parameter
$w_{M}(t) \equiv P_{M}(t)/\rho_{M}(t)=1/3$.

\begin{figure*}[p]
\begin{center}
\hspace*{0mm}
%%26may2026: relabel FIG12345-v1.eps --> FIG12345-v3eps
\includegraphics[width=0.55\textwidth]
{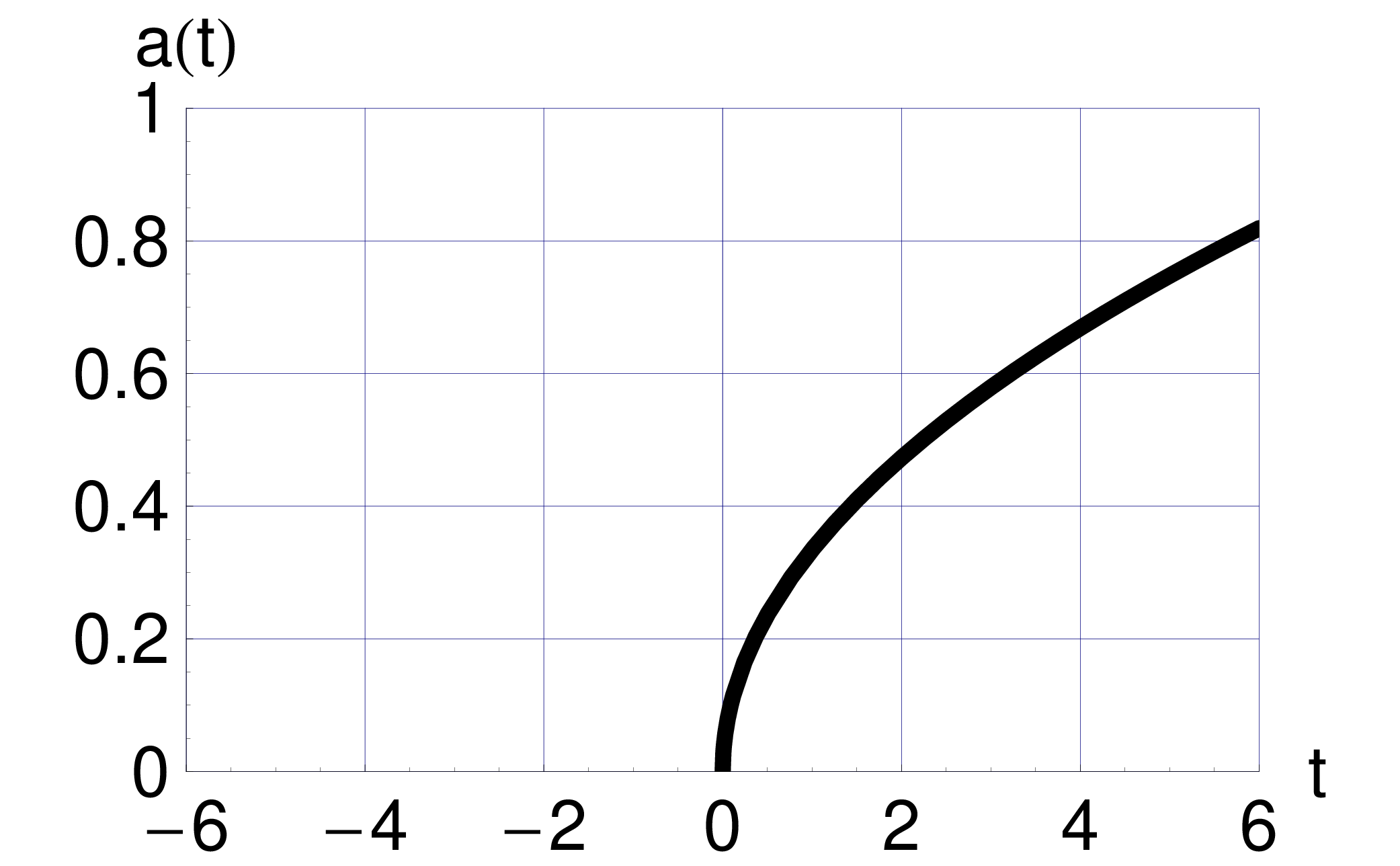}
%%{corfu2025-FIG1-v089.eps} %%29mar2026 new range
%%{corfu2025-FIG1-v012.eps}
%%{BB-as-spacetime-defect-FIG01-v4.eps}
%%{BB-as-spacetime-defect-FIG1-v012.eps}
%%corfu2025-FIGs.nb <=== NEW, on 29mar2026 with YO PC
\end{center}
\vspace*{8pt}
\caption{Cosmic scale factor $a(t)$ of the
Friedmann solution (\ref{eq:Friedmann-asol})
for equation-of-state parameter
$w_{M}=1/3$, with
$t_{0}=4\,\sqrt{5} \approx 8.944$.}
\protect\label{fig:a-FLRW}
\vspace*{0.1000mm}
\end{figure*}

The Friedmann solution~\eqref{eq:Friedmann-asol},
shown in Fig.~\ref{fig:a-FLRW}, has a
\textbf{Big Bang curvature singularity}:%
\beq
\lim_{t\to 0^{+}} a(t)\,\Big|_\text{F} =0\,,
\eeq
where the matter energy density $\rho_{M}$ and the
Kretschmann curvature scalar
$K\equiv R^{\mu\nu\rho\sigma}\,R_{\mu\nu\rho\sigma}$
\emph{diverge} at $t=0^{+}$. Hence,
the dynamics from the Einstein
equation \eqref{eq:Einstein-eq} has restricted
the range of the cosmic time coordinate $t$
to the open interval, here taken to be $(0,\,\infty)$.

Apparently, the ``journey-back-in-time''
by Friedmann et al. has ended up in a state
of infinite energy density and curvature.
This may hold more generally
(different matter content and/or less spatial symmetry),
as suggested by the Hawking--Penrose
singularity theorems~\cite{HawkingPenrose1970}
(see also the further remarks in
Sec.~\ref{sec:Defect-cosmology} below).

Now comes the question announced in the
section header and the surprising answer:
\begin{description}
  \item[Q1:]
Is it possible to \underline{evade}
the Friedmann curvature
singularity \underline{without}
introducing new particles,
new fields, or new physics?
  \item[A1:]
\underline{Yes}, and we will find that
the tamed Big Bang has multiple ``sides.''
\end{description}

%%\newpage%%tmp
\section{Defect cosmology and a follow-up question}
\label{sec:Defect-cosmology}

A new spatially-flat metric \textit{Ansatz}
has been proposed~\cite{Klinkhamer2019}:%
\bsubeqs\label{eq:RWK}
\beqa\label{eq:RWK-ds2}
\hspace*{0mm}
ds^{2}\,\Big|_\text{RWK}
%&\equiv&
%g_{\mu\nu}(x)\, dx^{\mu}\,dx^{\nu} \,
%\Big|_\text{RWK}
%\nonumber\\[1mm]
&=&
- \frac{t^{2}}{b^{2}+t^{2}}\,d t^{2}
+ a^{2}( t )\;
%\Big(dr^{2}+ r^{2}\,
%\Big[ d\theta^{2} + \sin^{2}\theta\, d\phi^{2} \Big]\Big)
%%\delta_{m n}\,dx^{m}\,dx^{n}
\Big[ \big(d x^{1}\big)^{2} +
\big(d x^{2}\big)^{2} +\big(d x^{3}\big)^{2} \Big]
\,,
\\[2mm]
\label{eq:RWK-b-positive}
\hspace*{0mm}
b &>& 0\,,
\\[2mm]
\hspace*{0mm}
a( t ) &>& 0\,,
\\[2mm]
\label{eq:RWK-ranges-coordinates}
\hspace*{0mm}
% t    &\in& (-\infty,\,\infty)\,,\\[2mm]
%x^{m} \in (-\infty,\,\infty)
% r    &\in& [0,\,\infty)\,,\quad
%\theta \in [0,\,\pi]\,,\quad
%\phi \in [0,\,2\pi)
t,\,x^{1},\,x^{2},\,x^{3} &\in& (-\infty,\,\infty)
\,.
\eeqa
\esubeqs
This is a \textbf{degenerate metric},
having a vanishing determinant at $t = 0$.
The $t = 0$ slice corresponds
to a three-dimensional \textbf{spacetime defect},
analogous to defects in an atomic crystal;
see, e.g., the research
paper~\cite{KlinkhamerSorba2014} and the two
reviews~\cite{{Klinkhamer2014-MPLAreview},Klinkhamer2019-JPCSreview}
for further discussion and references.
Instead of ``spacetime defects,'' it may also be
useful to think of ``weaving errors'' in the
fabric of spacetime.

\begin{figure*}[p]
\begin{center}
\hspace*{0mm}
\includegraphics[width=0.55\textwidth]
{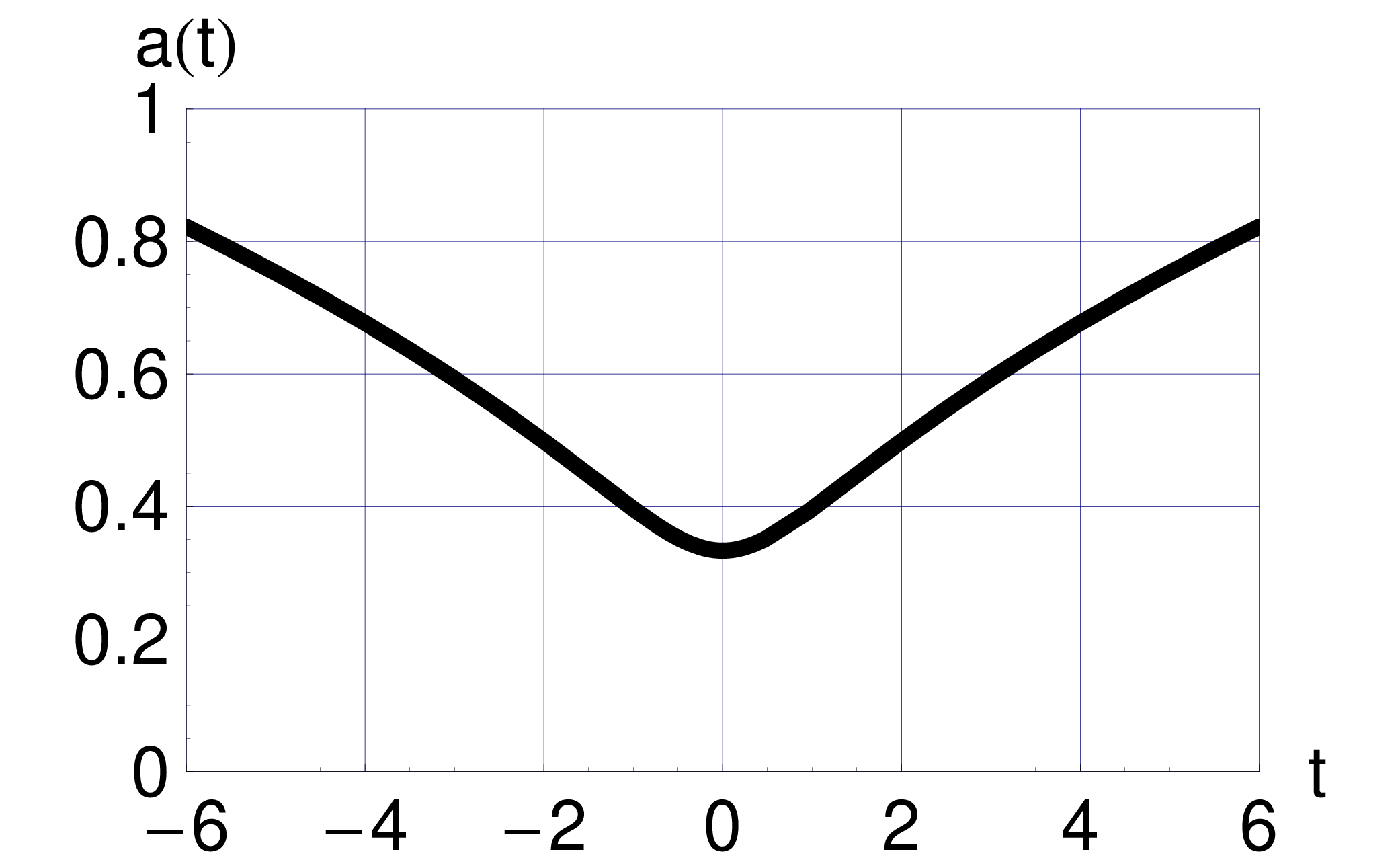}
%%{corfu2025-FIG2-v012.eps}
%%{BB-as-spacetime-defect-FIG02-v4.eps}
%%{BB-as-spacetime-defect-FIG2-v012.eps}
%{Proc-Cracow-Epiphany-2021-Klinkhamer-fig1-v1.eps}
%%{nonsingular-bouncing-cosmology-fig02-v5NEW-wMthird.eps}
%%nonsingular-bouncing-cosmology-2019.nb
%%corfu2025-FIGs.nb   <=== NEW
\end{center}
\vspace*{0pt}
\caption{Cosmic scale factor $a(t)$ of the
defect-cosmology solution
(\ref{eq:regularized-Friedmann-asol}) for
$w_{M}=1/3$, with $b=1$ and
$t_{0}=4\,\sqrt{5} \approx 8.944$.}
\protect\label{fig:a-FLRWK}
\vspace*{0.1000mm}
\end{figure*}

We now insert the metric (\ref{eq:RWK})
and the energy-momentum tensor of a
homogeneous relativistic perfect fluid
($w_{M}=1/3$)
into the Einstein gravitational field equation
\eqref{eq:Einstein-eq} and use
a ``continuous-extension''
procedure~\cite{Horowitz1991,Guenther2017}
for $t \to 0$
(see Sec.~\ref{subsec:ext-Einstein-eq}
for further discussion). This gives
the following \textbf{nonsingular solution}
(cf. Fig.~\ref{fig:a-FLRWK}):%
\bsubeqs\label{eq:regularized-Friedmann-asol-rhoMsol}
\beqa\label{eq:regularized-Friedmann-asol}
a(t)\,\Big|_\text{K}^{(w_{M}=1/3)}
&=&
\sqrt[4]{\big(b^{2}+t^{2}\big)\big/
\big(b^{2}+t_{0}^{2}\big)}\,,
%%\eeqa
\\[4mm]
%%\beqa
\label{eq:regularized-Friedmann-rhoMsol}
\rho_{M}(t)\,\Big|_\text{K}^{(w_{M}=1/3)} &=&
3\,P_{M}(t)\,\Big|_\text{K}^{(w_{M}=1/3)}
%%=\rho_{M,0}/a^{4}(t) \,\Big|_\text{K}^{(w_{M}=1/3)}
=
\rho_{M,0}\,
\big(b^{2}+t_{0}^{2}\big)\big/\big(b^{2}+t^{2}\big) \,,
\eeqa
\esubeqs
where the energy density $\rho_{M}$ and the
Kretschmann curvature scalar $K$
are \emph{finite} at $t=0$, respectively of
order $E_\text{Planck}^{2}/b^{2}$ and $1/b^{4}$.
Still, certain quantities are discontinuous
at the $t=0$ defect~\cite{Battista2021,Wang2021},
in line with the Hawking--Penrose singularity
theorems~\cite{HawkingPenrose1970}
as reviewed in Sec.~4.5 of
Ref.~\cite{Klinkhamer2025-MPLAreview}.

\begin{figure*}[p]
\begin{center}
\hspace*{0mm}
\includegraphics[width=0.55\textwidth]
{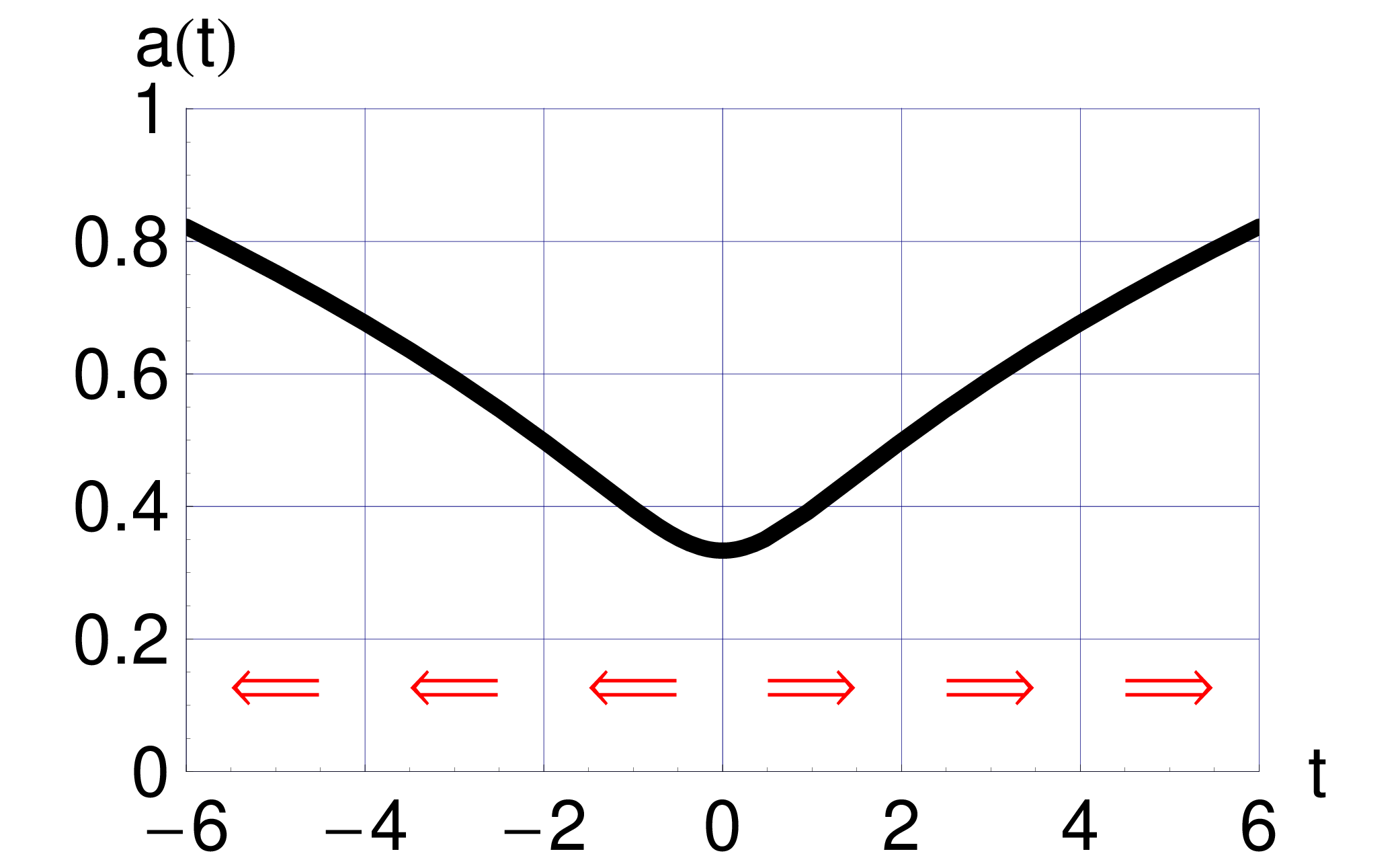}
%%corfu2025-FIGs.nb   <=== NEW
\end{center}
\vspace*{8pt}
\caption{Defect cosmology: Pair-creation scenario.
The small double arrows (in red)
indicate the direction of the physical,
thermodynamic time $\mathcal{T}=|t|$;
see Sec.~\ref{sec:Defect-cosmology}
for further discussion.}
\protect\label{fig:a-FLRWK-ARROWS}
\vspace*{0.1000mm}
\end{figure*}

\begin{figure*}[p]
\centerline{
\includegraphics[width=0.3\textwidth]{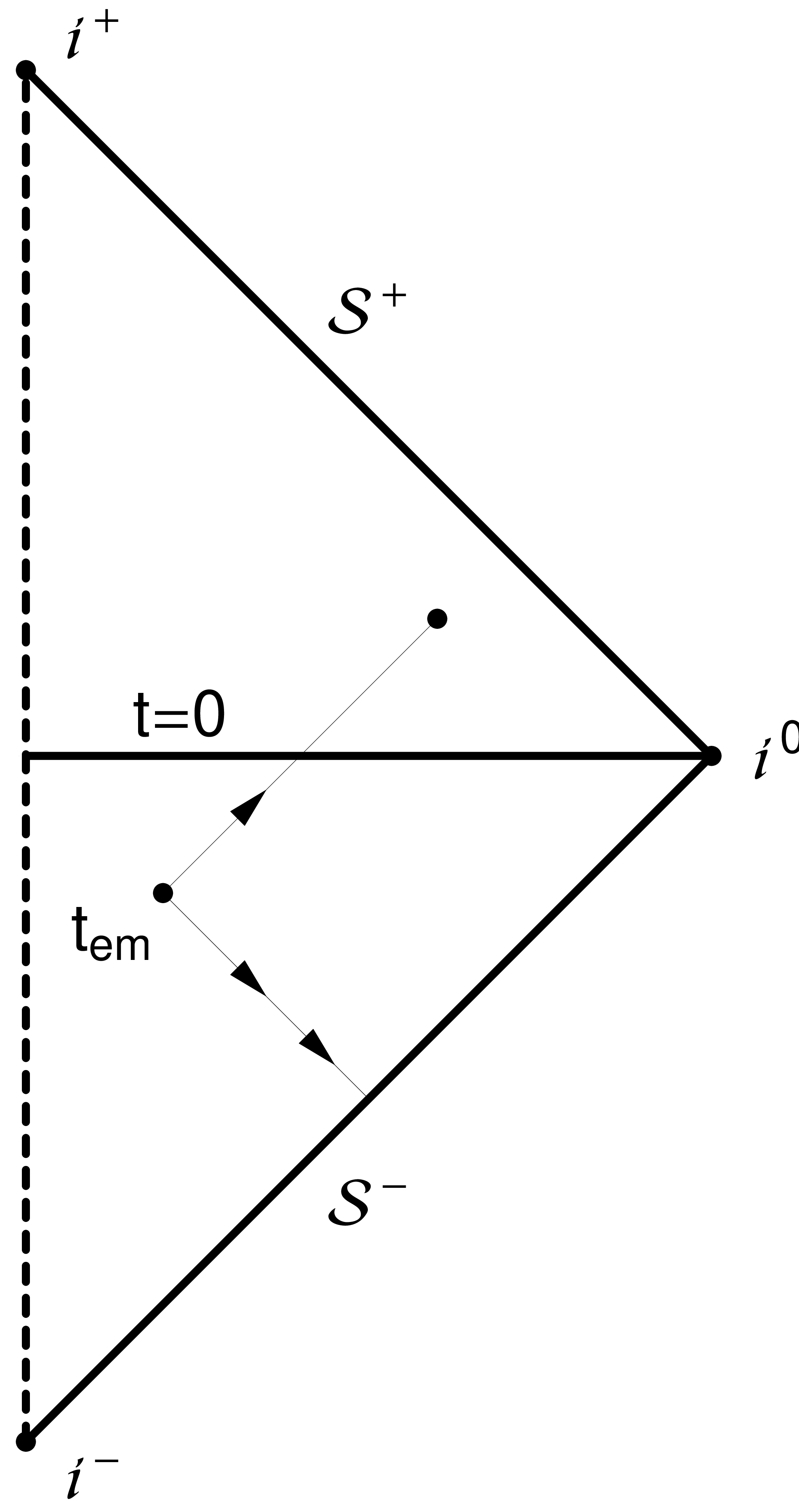}}
%%{BB-as-spacetime-defect-FIG03-v4.eps}
%%{BB-as-spacetime-defect-FIG4-v270.eps}
\vspace*{-0.5\baselineskip}
\caption{Defect cosmology:
Penrose conformal diagram
with the same notation as in Fig.~15(ii) of
Ref.~\cite{HawkingEllis1973}.
A $t$-retarded gravitational wave with emission time
$t_\text{em}<0$  and observation time $t_\text{obs}>0$
is shown as the single-arrowed curve going diagonally up
and a $t$-advanced gravitational
wave emitted at $t_\text{em}<0$
as the double-arrowed curve going diagonally down;
see Sec.~\ref{sec:Defect-cosmology}
for further explanations.}
\protect\label{fig:FLRWK-conf-diagram}
\vspace*{0.1000mm}
\end{figure*}

\begin{figure*}[p]
\begin{center}
\hspace*{0mm}
\includegraphics[width=0.75\textwidth]
{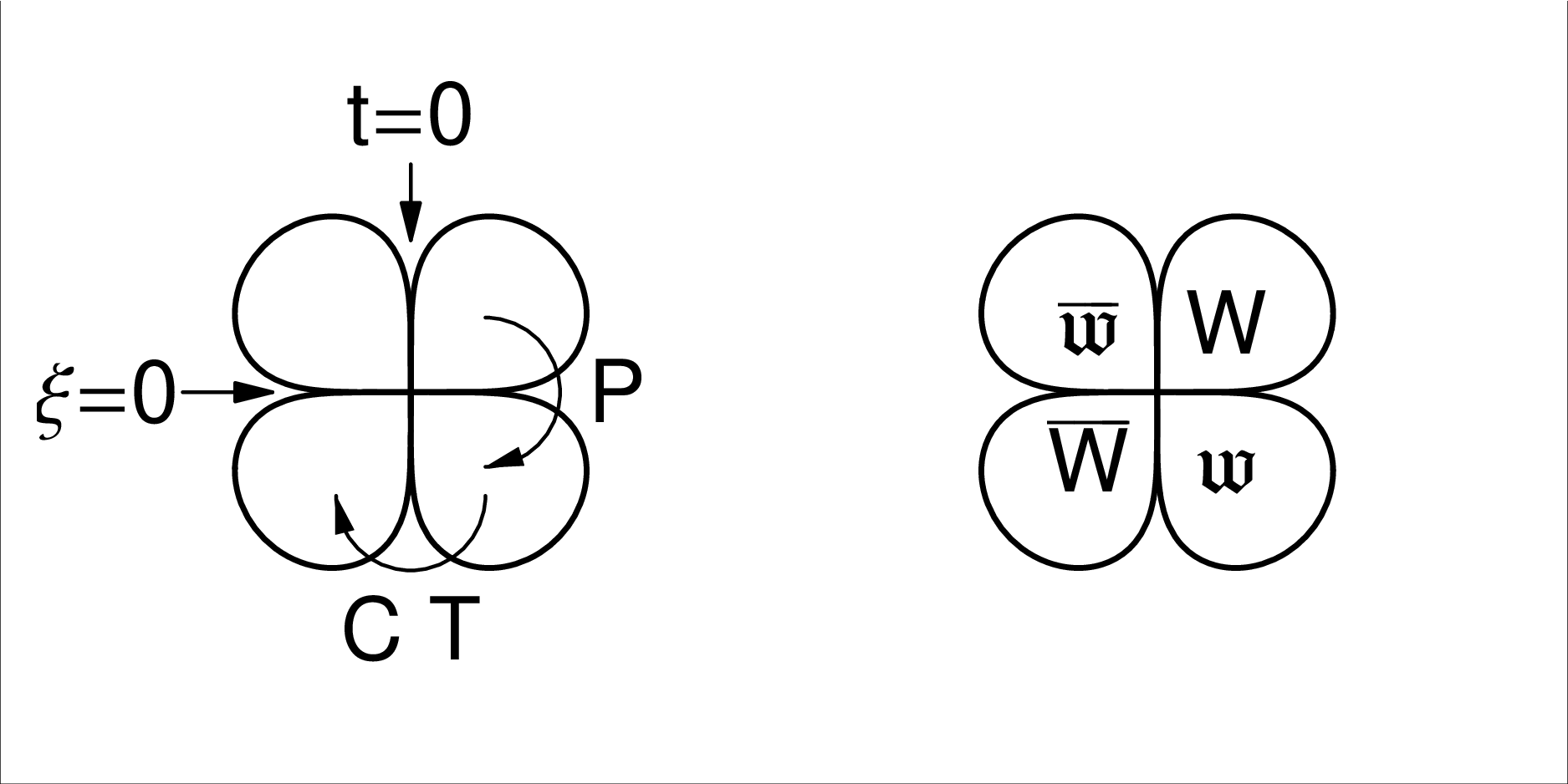}
%%{corfu2025-FIG5-v092.eps}
%%corfu2025-FIGs.nb   <=== NEW
\end{center}
\vspace*{0pt}
\caption{Two sketches of the
four-leaf-clover universe.
Left: spacetime defects at $t=0$ and $\xi=0$, with
exemplary P and CT transformations.
Right: two pairs of CPT-conjugated worlds
\big($W$--$\overline{W}$
and $\mathfrak{w}$--$\overline{\mathfrak{w}}$\big).
See Sec.~\ref{sec:Defect-cosmology-FLC-universe}
for further explanations and references.}
\protect\label{fig:clover}
\vspace*{0.1000mm}
\end{figure*}

There are several cosmological scenarios.
The two main ones are:
\begin{itemize}
  \item
The \textbf{defect bounce scenario},
with a contracting universe for $t<0$
(pre-bounce phase) and
an expanding universe for $t>0$
(post-bounce phase),
with a bounce at $t=0$ where
 $\mathrm{d} a/\mathrm{d} t = 0$.
  \item
The \textbf{defect pair-creation scenario},
where a spacetime defect
is created at $t=0$ with basically identical
worlds for $t<0$ and $t>0$.
\end{itemize}

%%\newpage%%tmp
The first scenario~\cite{KlinkhamerWang2019-cosm,%
KlinkhamerWang2020-pert},  with
the coordinate time $t$ interpreted as the
``physical'' time, is more or less
standard.
But, now, there is no
need for exotic matter to make the bounce
[turning part of the concave curve $a(t)$
into a convex segment around $t=0$;
cf. Fig.~\ref{fig:a-FLRWK}],
as the exoticness is already carried by the metric.
This first scenario leaves, however,
the origin of the
spacetime defect completely open,
as well as the mechanism to produce
the appropriate initial
boundary conditions at $t \ll -b$.

In the second scenario (cf. Sec.~6 of the
review~\cite{Klinkhamer2025-MPLAreview}
for a brief discussion with further references),
we assume smoothness at $t=0$,
possibly with small matter perturbations
present. These
matter perturbations grow as a power of $t^{2}$,
hence symmetrically in coordinate time $t$;
details can be found in
Ref.~\cite{KlinkhamerWang2020-pert}.
Then, we can interpret the physical time as
the ``thermodynamic''
time $\mathcal{T}\equiv |t|$
for which matter density perturbations grow
(and entropy increases).
We see that \emph{both} worlds
($t<0$ and $t>0$) are \emph{expanding}
with respect to $\mathcal{T}\equiv |t|$
(cf. Fig.~\ref{fig:a-FLRWK-ARROWS}).
Furthermore, with $\mathcal{T}$ the
physical time, we have that gravitational
waves from \emph{standard} emission processes
will be $\mathcal{T}$-retarded waves.
These $\mathcal{T}$-retarded gravitational waves
then correspond, for $\mathcal{T}\equiv |t|$,
to $t$-retarded waves in the $t>0$ phase
and $t$-advanced waves in the $t<0$ phase;
see Sec.~5 in Ref.~\cite{Klinkhamer2025-MPLAreview}
for a detailed discussion.

The conformal diagram of
defect cosmology is shown in
Fig.~\ref{fig:FLRWK-conf-diagram}.
As mentioned in the previous paragraph,
%standard emission processes
%operative in the $t<0$ phase
a gravitational wave generated
in the $t<0$ phase
propagates as the single-arrowed curve
according to the first scenario
(a $t$-retarded wave from $\mathcal{T} \equiv t$)
and as the double-arrowed curve
according to the second scenario
(a $t$-advanced wave from $\mathcal{T}\equiv |t|$).

In the defect pair-creation scenario,
we apparently have two expanding worlds
(that is, expanding with respect to
the thermodynamic
time $\mathcal{T}\equiv |t|$;
cf. Fig.~\ref{fig:a-FLRWK-ARROWS}).
These two worlds are CT-reversed copies of
each other but \emph{not} parity-reversed copies.

Now comes the follow-up question
announced in the section header and
the answer:
\begin{description}
  \item[Q2:]
Is it possible to change the metric
\textit{Ansatz}, so as to get a pair
of \underline{CPT-conjugated worlds}
(as discussed previously by
Sakharov~\cite{Sakharov1967} and
Boyle, Finn, and
Turok~\cite{BoyleFinnTurok2018})?
  \item[A2:]
Yes, the \underline{four-leaf-clover universe},
which has, in fact, \underline{two} pairs
of CPT-conjugated
worlds, as will be discussed
in the next section.
\end{description}

%%\newpage%%tmp

\section{Defect cosmology: Four-leaf-clover universe}
\label{sec:Defect-cosmology-FLC-universe}

The basic idea is to
combine a cosmological
defect~\cite{Klinkhamer2019}
at $t=0$ and a wormhole
defect~\cite{Klinkhamer2023a,Klinkhamer2023b}
at $\xi=0$, where $\xi$ is a quasi-radial coordinate.
The new four-leaf-clover (f-l-c)
metric then reads~\cite{Klinkhamer2026}:%
\bsubeqs\label{eq:f-l-c}
\beqa\label{eq:f-l-c-ds2}
\hspace*{0mm}
ds^{2}\,\Big|_\text{f-l-c}
&\equiv&
%g_{\mu\nu}(x)\, dx^{\mu}\,dx^{\nu} \,
%\Big|_\text{f-l-c}=
- \frac{t^{2}}{b^{2}+t^{2}}\,d t^{2}
\nonumber\\[1mm]
&&
+ a^{2}( t )
\;\Bigg(\frac{\xi^{2}}{\lambda^{2} + \xi^{2}}   d\xi^{2}
+ \left(\lambda^{2} + \xi^{2}\right)\,
  \Big[ d\theta^{2} + \sin^{2}\theta\, d\phi^{2} \Big]\Bigg)
%%\delta_{m n}\,dx^{m}\,dx^{n}
\,,
\\[2mm]
\label{eq:f-l-c-b-positive}
\hspace*{0mm}
b\,,\lambda &>& 0\,,
\\[2mm]
\label{eq:f-l-c-a-positive-even}
\hspace*{0mm}
a( t ) &>& 0\,,\quad
a(-t ) = a( t )\,,
\\[2mm]
\label{eq:f-l-c-ranges-coordinates}
\hspace*{0mm}
 t\,,\xi    &\in& (-\infty,\,\infty)\,,\quad
%%x^{m} \in (-\infty,\,\infty)
\theta \in [0,\,\pi]\,,\quad
\phi \in [0,\,2\pi)\,,
\eeqa
\esubeqs
where the spatial coordinates
$\{\xi,\, \theta,\,\phi\}$ provide two copies
(one for $\xi <0$ and another for $\xi >0$)
of the standard spherical coordinates of $\mathbb{R}^{3}$,
making for a double cover of Euclidean 3-space
(see, e.g.,
Fig.~A.1 of Ref.~\cite{Klinkhamer2025-MPLAreview}
for a sketch of a wormhole connecting two
3-spaces).
With a homogeneous relativistic perfect fluid,
we get the same
ordinary differential equations (ODEs)
as before and, therefore, the same solution
(\ref{eq:regularized-Friedmann-asol-rhoMsol}),
but now in \emph{two} copies,
one for $\xi <0$ and another for $\xi >0$.
%for ODEs, see four-leaf-clover-universe-2025.nb

The corresponding dual basis (1-forms defined
by $e^{a} \equiv e^{a}_{\phantom{z}\mu}\,\text{d}x^{\mu}$
in terms of the tetrad $e^{a}_{\phantom{z}\mu}$)
reads~\cite{Klinkhamer2026}:%
\begin{subequations}\label{eq:f-l-c-tetrad}
\begin{eqnarray}
\label{eq:f-l-c-tetrad-a=0}
e^{0}\,\Big|_\text{f-l-c}
&=&  \frac{t}{\sqrt{b^{2} + t^{2}}}\; \text{d}t\,,
\;\;\qquad\qquad \leftarrow \text{T-odd}
\\[1mm]
\label{eq:f-l-c-tetrad-a=1}
e^{1}\,\Big|_\text{f-l-c}
&=& a( t )\, \frac{\xi}{\sqrt{\lambda^{2} + \xi^{2}}}\; \text{d}\xi\,,
\qquad \leftarrow \text{P-odd}
\\[1mm]
e^{2}\,\Big|_\text{f-l-c}
&=& a( t )\, \sqrt{\lambda^{2} + \xi^{2}}\; \text{d}\theta\,,
\\[1mm]
e^{3}\,\Big|_\text{f-l-c}
&=& a( t )\, \sqrt{\lambda^{2} + \xi^{2}}\;\sin\theta  \;  \text{d}\phi\,.
\end{eqnarray}
\end{subequations}
In these tetrads,
the time-reversal transformation (T)
and parity-reversal transformation (P)
are ``hard-wired,''
as shown by \eqref{eq:f-l-c-tetrad-a=0}
and \eqref{eq:f-l-c-tetrad-a=1}, respectively.
Note that different numerators $|t|$ and $|\xi|$ in
these two particular tetrads would remove
the odd T and P reflection properties,
but would also introduce (unwanted)
delta-function singularities of the
curvature~\cite{Klinkhamer2025-MPLAreview}.

%%\newpage%%tmp
Heuristically, the
charge-conjugation transformation (C) is
expected to join the T transformation,
according to the Stueckelberg--Feynman
interpretation of antiparticles as
particles running backward in time
(see, in particular, Footnote~7
in Ref.~\cite{Feynman1949} and also
Figure~2 there).
But the real explanation of the
appearance of the C transformation
requires a better understanding of the
pair-creation \emph{mechanism} and,
hence, of the underlying fundamental theory.
For the moment, we can only assume
the validity of an \emph{effective} CPT theorem
(as discussed in the third paragraph of Sec.~V
in Ref.~\cite{Klinkhamer2026}), so that
antiparticles would appear in the $t<0$ phase
instead of particles in the $t>0$ phase,
given the parity and time reversals of the
tetrads~\eqref{eq:f-l-c-tetrad}.

Further details on the model and its phenomenology
(e.g., classical communication between the four
worlds of Fig.~\ref{fig:clover})
can be found in Ref.~\cite{Klinkhamer2026}.

%%\newpage%%tmp
\section{Behavior at the t=0 defect hypersurface}
%%pdflatex does not like:s  $\mathbf{t=0}$
\label{sec:Behavior-at-tis0}

\subsection{Modified Friedmann equations}
\label{subsec:mod-Friedmann-eqs}

In Secs.~\ref{sec:Friedmann-cosmology}
and \ref{sec:Defect-cosmology},
we have not shown explicitly the standard
and modified Friedmann equations that give, respectively,
the singular solution~\eqref{eq:Friedmann-solution} and
the nonsingular solution~\eqref{eq:regularized-Friedmann-asol-rhoMsol},
but these two sets of ODEs
can be found in, e.g.,
the review~\cite{Klinkhamer2025-MPLAreview}.

Here, we will discuss in some detail how the
nonsingular
solution~\eqref{eq:regularized-Friedmann-asol-rhoMsol}
comes about, starting from
the metric \textit{Ansatz}~\eqref{eq:RWK}.
Let us begin with some ``close reading''
of these modified Friedmann equations, which
are explicitly:
\bsubeqs\label{eq:mod-Feqs}
\beqa
\label{eq:mod-1stFeq}
\hspace*{-0mm}&&
\left[1+ \frac{b^{2}}{t^{2}}\,\right]\,
\left( \frac{\dot{a}}{a}\right)^{2}
= \frac{8\pi G}{3}\,\rho_{M}\,,
\\[1mm]
\label{eq:mod-2ndFeq}
\hspace*{-0mm}&&
\left[1+\frac{b^{2}}{t^{2}}\,\right]\,
\left(\frac{\ddot{a}}{a}+
\frac{1}{2}\,\left( \frac{\dot{a}}{a}\right)^{2}
\right)
-\frac{b^{2}}{t^{3}}\,\frac{\dot{a}}{a}
=
-4\pi G\,P_{M}\,,
\\[1mm]
\label{eq:mod-Feq-rhoMprimeeq}
\hspace*{-0mm}&&
\dot{\rho}_{M}+ 3\;\frac{\dot{a}}{a}\;
\Big(\rho_{M}+P_{M}\Big) =0\,,
\\[1mm]
\label{eq:mod-Feq-EOS}
\hspace*{-0mm}&&
P_{M} = P_{M} \big(\rho_{M}\big)\,,
\eeqa
\esubeqs
where the overdot stands for differentiation
with respect to $t$.
Equation \eqref{eq:mod-Feq-rhoMprimeeq} corresponds to
energy conservation (discussed further
in App.~\ref{subapp:Energy-mom-conserv})
and Eq.~\eqref{eq:mod-Feq-EOS} to the
equation of state of the perfect fluid.
For a generic function $a(t)$ with $a(t)>0$
by assumption, the first two ODEs are
singular at $t=0$
because of the explicit factors $1/t^{2}$
and $1/t^{3}$.

Now consider a function
with bounce behavior near the
coordinate origin $t=0$,
\begin{subequations}\label{eq:a-bounce}
\begin{eqnarray}
\label{eq:a-bounce-terms}
a_\text{bounce}(t)
&=&
1 + a_{2} \,(t/b)^{2}
+ a_{4} \,(t/b)^{4}
+ O\left((t/b)^{6}\right)\,, \;\;
\text{for}\;\;t\sim 0 \,,
\\[1mm]
\label{eq:a-bounce-a2pos}
a_{2} &>& 0\,,
\end{eqnarray}
\end{subequations}
where the function $a(t)$ has been normalized to unity
at the origin, $a(0)=1$.
[Obviously, the bounce behavior is a property of
the full solution \eqref{eq:regularized-Friedmann-asol}
as shown in Fig.~\ref{fig:a-FLRWK}.]
Then, the ``dangerous'' terms in the
ODEs~\eqref{eq:mod-Feqs}
are made finite as follows:
\begin{subequations}\label{eq:finitized-terms}
\begin{eqnarray}
\label{eq:finitized-term1}
\frac{b^{2}}{t^{2}}\,
\left(\frac{\dot{a}}{a}\right)^{2}
&\to&  4 \, a_{2}^{2} \,b^{-2}
+ O\left( t^{2}/b^{4}\right)  \,,
\\[1mm]
\label{eq:finitized-term2}
\frac{b^{2}}{t^{2}}\,
\left(\frac{\ddot{a}}{a}\right)
-\frac{b^{2}}{t^{3}}\,\frac{\dot{a}}{a}
 &\to&
 + 2\, a_{2}/t^{2} - 2\, a_{2}/t^{2}
 + O\left( b^{-2}\right)
 \nonumber\\[1mm]
 &=&
 8 \, a_{4} \,b^{-2}
+ O\left( t^{2}/b^{4}\right) \,,
\end{eqnarray}
\end{subequations}
where, for the last equation,
both divergent contributions
cancel precisely as shown
on the first line, leaving the
finite terms shown on the second line.
Similar results hold for the curvature scalars
$R$ and $K$, while
$\rho_{M}(t) \propto 1/a^{4}(t)$
is manifestly finite at $t=0$, as $a(0)$ is
nonzero according to \eqref{eq:a-bounce-terms}.

%%\newpage%%tmp
There is an important point we need to make.
The cancellation of the two divergent terms
shown in the first line
of \eqref{eq:finitized-term2}
holds, strictly speaking, only for $t\ne 0$,
because these $1/t^{2}$ terms
are ill-defined at $t= 0$.
For this reason, we had to use a
continuous-extension procedure ($t \to 0$)
for the derivation of the
ODEs~\eqref{eq:mod-Feqs}.

%%\newpage%%tmp
\subsection{Extended Einstein equation}
\label{subsec:ext-Einstein-eq}

The next question we like to address is
how the ODEs~\eqref{eq:mod-Feqs} are derived
in the first place. As mentioned
in the lines
above~\eqref{eq:regularized-Friedmann-asol-rhoMsol},
this derivation relies on the procedure of
continuous extension;
see Ref~\cite{Horowitz1991}
and Sec.~3.3.1 of Ref.~\cite{Guenther2017}.
In this logic, the spacetime points of the defect,
at $t=0$ for the metric~\eqref{eq:RWK},
are treated differently than the $t\ne 0$
spacetime points
of the ambient spacetime in which the defect
is embedded.

But there is an alternative approach where
\underline{all} spacetime
points are, in principle, treated \underline{equally}.
The approach postulates the relevance
of a \underline{different} gravitational field
equation. This idea has first been suggested by
Einstein and Rosen~\cite{EinsteinRosen1935} and
has been discussed later by Horowitz~\cite{Horowitz1991}
in particular, followed by the
present author in Sec.~4.4 of Ref.~\cite{Klinkhamer2025-MPLAreview}
(see also Ref.~\cite{Dimaschko2025}
for a recent pedagogical discussion
in the context of wormholes).

In short, the idea
is to replace the standard
Einstein field equation~\eqref{eq:Einstein-eq}
by an extended version,
which is obtained by multiplying the whole standard equation
by the square of the metric determinant:%
\begin{subequations}
\label{eq:extended-Einstein-eq}
\begin{eqnarray}
g^{2}\,R_{\mu\nu}
&=&
8\pi G\;g^{2}\,
\left(T_{\mu\nu}^\text{\,(M)}
- \frac12\, g_{\mu\nu}\,T^\text{\,(M)}\right) \,,
\\[1mm]
g &\equiv& \det (g_{\mu\nu}) \,.
\end{eqnarray}
\end{subequations}
For the vacuum case, this is precisely the postulated
 equation $g^{2}\,R_{\mu\nu}=0$
discussed by Einstein and Rosen~\cite{EinsteinRosen1935}
(and later in Refs.~\cite{Horowitz1991}
and \cite{Klinkhamer2025-MPLAreview,Dimaschko2025}).
The crucial observation is that
$g^{2}\,R_{\mu\nu}  \equiv W_{\mu\nu}$
turns out to be a continuous function of the metric
$g_{\alpha\beta}$ and its first two derivatives
but \emph{not} of the inverse metric $g^{\alpha\beta}$;
see App.~\ref{subapp:LHS-Sketch-proof}
for a brief explanation.
For this reason, it is possible that certain
degenerate metrics, with ill-defined inverses,
provide a solution of the extended equation
\eqref{eq:extended-Einstein-eq}
at all spacetime points
(presupposing a well-behaved right-hand side for
the perfect-fluid energy-momentum tensor;
see App.~\ref{subapp:RHS} for further discussion).
The defect-cosmology metric~\eqref{eq:RWK}
is an example of such a degenerate metric.

%%\newpage%%tmp
At the risk of belaboring the obvious:
any metric $\overline{g}_{\mu\nu}$
that solves the standard Einstein
equation~\eqref{eq:Einstein-eq}
at all spacetime points
also solves \eqref{eq:extended-Einstein-eq}
at all spacetime points,
but the opposite need not be true.
The defect-cosmology metric~\eqref{eq:RWK}
indeed solves \eqref{eq:extended-Einstein-eq}
at all spacetime points but
not \eqref{eq:Einstein-eq}, as that last equation
becomes ill-defined
at $t=0$ (certain components of the affine connection
$\Gamma^{\lambda}_{\mu\nu}$ diverge for $t=0$,
cf. Refs.~\cite{Wang2021,Battista2021}), which
explains the need for the
continuous-extension procedure $t \to 0$
in a derivation based on the
standard Einstein equation.

Expanding on the last remark,
it may be worthwhile to
reconsider the modified Friedmann
ODEs~\eqref{eq:mod-Feqs}, now
derived from the extended equation
\eqref{eq:extended-Einstein-eq}.
The resulting ODEs
would be equal to those of \eqref{eq:mod-Feqs}
multiplied by an overall factor
\begin{equation}
\label{eq:g2}
g^{2}\,\Big|_\text{RWK} =
\frac{t^{4}}{(b^{2}+t^{2})^{2}}\;a^{12}(t) \,,
\end{equation}
for the metric~\eqref{eq:RWK} in terms of Cartesian
spatial coordinates.
Neglecting nonsingular factors $(b^{2}+t^{2})^{-2}$
and $a^{12}(t)$ in this $g^{2}$ factor
and again considering
the bounce behavior \eqref{eq:a-bounce} of
the cosmic scale factor, the troublesome
terms of \eqref{eq:finitized-terms} now become
\begin{subequations}\label{eq:new-finitized-terms}
\begin{eqnarray}
\label{eq:new-finitized-term1}
 b^{2}\,t^{2}\,
\left(\frac{\dot{a}}{a}\right)^{2}
&\to&  4 \, a_{2}^{2} \,b^{-2}\, t^{4}
+ O\left( t^{6}/b^{4}\right)  \,,
\\[0mm]
\label{eq:new-finitized-term2}
b^{2}\,t^{2}\,
\left(\frac{\ddot{a}}{a}\right)
-b^{2} \,t\,\frac{\dot{a}}{a}
 &\to&
 + 2\, a_{2}\, t^{2} - 2\, a_{2}\, t^{2}
 + O\left( b^{-2}\, t^{4}\right)
 \nonumber\\[0mm]
 &=&
 8 \, a_{4}  \, t^{4}  \,b^{-2}
+ O\left( t^{6}/b^{4}\right) \,,
\end{eqnarray}
\end{subequations}
where the cancelation shown on the first line
of the last equation holds for
\emph{any} value of $t$, including the
value $t=0$.

%%\newpage%%tmp
\subsection{Discussion}
\label{subsec:Discussion}

The crucial question is whether or not
\eqref{eq:extended-Einstein-eq}
holds true and, if so,
what the origin is of this extended version
of Einstein's field equation.
At this moment, we do not know the answer.

It is possible to think of higher-dimensional
effects or, more specifically, dilaton-type
effects~\cite{GreenSchwarzWitten1987,Polchinski1998}.
%% cf. 26D Eq.(3.4.58) in GSW ; 10D Eq.(16.4.11) in Pol
With the dimensionless dilaton field $\phi(x)$,
there is then
an overall factor $\exp[-2\phi(x)]$
in the integrand of the
effective 4-dimensional action,
\beq
S_\text{eff}=
\int \,d^4x\, \sqrt{-g}\;
e^{-2\phi}\,
\left[
\frac{1}{2\kappa^{2}}\,
\left(
R-\lambda
+4 g^{\mu\nu} \partial_{\mu}\phi\,\partial_{\nu}\phi
\right)
+\mathcal{L}_{M}
\right]\,,
\eeq
where $\lambda$ is a cosmological constant
with dimension of inverse length squared
and $\mathcal{L}_{M}$ is the Lagrange density
of the  matter.
By variation of the metric, we then get
the standard Einstein
equation~\eqref{eq:Einstein-eq}
multiplied by the scalar factor $\exp[-2\phi(x)]$.
Hypothetically, it could be that
the dilaton expectation value
$\langle\phi\rangle$ gives
an overall factor $\exp[-2\langle\phi\rangle]$
proportional to a
scalar $\Omega\,g^{2}$ for a constant
background scalar density $\Omega$ of weight $+4$
(i.e., the opposite of the $g^{2}$
weight~\cite{Weinberg1972}).
Assuming the existence of a
mechanism to generate $\Omega \ne 0$
and neglecting the constant $\lambda$ and
the dilaton derivative term in the action,
we would then obtain the extended Einstein
field equation~\eqref{eq:extended-Einstein-eq}.

Anyway, all kinds of speculation are possible
but they remain unconvincing
without some form of indirect evidence.

%%\newpage%%tmp
\section{Conclusions}
\label{sec:Conclusions}

The three main results reviewed in this
contribution are as follows,
starting with the most important one:
\begin{enumerate}
  \item
The Friedmann curvature singularity
can be evaded if we
consider a \underline{degenerate}
metric~\cite{Klinkhamer2019}.
  \item
The resulting ``tamed'' Big Bang has
\underline{multiple sides},
corresponding to our world and
others~\cite{KlinkhamerWang2019-cosm,%
KlinkhamerWang2020-pert}.
  \item
\underline{Classical communication} between
the $t<0$ and $t>0$
worlds may be difficult but perhaps not
impossible~\cite{Klinkhamer2025-MPLAreview}.
\end{enumerate}
In addition, we have,
in  Sec.~\ref{sec:Behavior-at-tis0},
carefully considered
the behavior of the nonsingular solution
at the $t=0$ defect hypersurface and recalled
the suggestion~\cite{EinsteinRosen1935}
that the extended Einstein
equation~\eqref{eq:extended-Einstein-eq}
could be relevant (a more technical
discussion is relegated to
App.~\ref{app:Additional-remarks-ext-Einstein-eq}).

The extended Einstein equation can then
be said to \underline{predict} a new effect,
namely the multiple sides of a tamed Big Bang
(interpreted as a spacetime defect).
But there could also be another new
effect appearing as the phenomenon
of ``antigravity,''
which will be reviewed in
App.~\ref{subapp:Another-effect-Antigravity}.

Still, the outstanding problem is to
\emph{derive} this extended Einstein
equation or, at least, to obtain an argument
for the appearance of an overall factor
proportional to a power of the metric determinant.

\acknowledgments
It is a pleasure to thank,
first, Zi-Liang Wang and Emmanuele Battista
for useful discussions over the years
and, second, Emmanuel Saridakis
and colleagues,  %%\emph{et al.}
for organizing the
``Workshop on Tensions in Cosmology''
(September 2--8)
at the Corfu Summer Institute 2025.

%%FRK: section-wise equation numbering for Apps!!!
\setcounter{equation}{0}
%%taken from: Rajesh [rajesh@wspc.com.sg] Monday, September 23, 2013 09:27
\renewcommand{\theequation}{\thesection.\arabic{equation}}

%%\newpage%%tmp
\appendix

\section{Additional remarks on the extended Einstein equation}
\label{app:Additional-remarks-ext-Einstein-eq}

In this appendix, we present some remarks to clarify the
interpretation of the extended Einstein
equation~\eqref{eq:extended-Einstein-eq} introduced in
Sec.~\ref{subsec:ext-Einstein-eq}, first
for the left-hand side (LHS) and then
for the right-hand side (RHS).

For ease of discussion, we repeat this
extended equation with minor simplifications and
generalizations:
\begin{equation}
\label{eq:ext-Einstein-eq-k=2}
g^{k}\,R_{\mu\nu}= 8\pi G\;g^{k}\,
\left(T_{\mu\nu}
- \frac12\, g_{\mu\nu}\,g^{\kappa\lambda}\,T_{\kappa\lambda}\right)\,,
\qquad   k=2\,,
\end{equation}
where we recall that $g$ is defined
as the determinant of the metric $g_{\mu\nu}$
and the Ricci tensor $R_{\mu\nu}$
as the contraction of the Riemann
tensor,
$R_{\mu\nu} \equiv R^{\lambda}_{\:\:\mu\lambda\nu}$.
In App.~\ref{subapp:LHS-Sketch-proof},
the choice $k=2$ will be seen to
suffice for the left-hand side of
\eqref{eq:ext-Einstein-eq-k=2},
but larger $k$ values might be
required on the right-hand side
for certain matter components,
as will be discussed at the
end of App.~\ref{subapp:RHS}.
For the moment, we just
keep $k=2$ as stated in
\eqref{eq:ext-Einstein-eq-k=2}.

\subsection{LHS: Sketch of a proof}
\label{subapp:LHS-Sketch-proof}

It is, of course, perfectly possible to evaluate the tensor
density $g^{2}\,R_{\mu\nu}$ directly.
But, here, we would like to
\emph{understand} how it comes that
the quantity $g^{2}\,R_{\mu\nu}$
does not involve the inverse metric $g^{\mu\nu}$,
but only the metric $g_{\mu\nu}$ and its derivatives.

We start by recalling the expression
for the Ricci tensor $R_{\mu\nu}$
already multiplied by $g^{2}$:
\begin{equation}
\label{eq:g2Rmunu-explicit} %%checked W1972  OK
g^{2}\,R_{\mu\nu} =
 g^{2}\,\partial_{\nu}\Gamma_{\mu \lambda }^{\lambda}
-g^{2}\,\partial_{\lambda}\Gamma_{\mu \nu}^{\lambda}
+\left(g\,\Gamma_{\mu \lambda}^{\kappa} \right)
\left(g\,\Gamma_{\nu \kappa}^{\lambda}\right)
-\left(g\,\Gamma_{\mu \nu}^{\kappa}\right)
\left(g\,\Gamma_{\lambda \kappa}^{\lambda}\right)
\,,
\end{equation}
defined in terms of the Christoffel symbol
\begin{equation}
\label{eq:Christoffel-def}  %%checked W1972  OK
\Gamma^{\lambda}_{\mu\nu}
=
\frac{1}{2}\,g^{\lambda \kappa}
\Big(\partial_{\mu} g_{\kappa\nu}
+\partial_{\nu} g_{\kappa\mu}
-\partial_{\kappa} g_{\mu \nu}\Big)\,,
\end{equation}
which displays the appearance of the inverse
metric $g^{\lambda \kappa}$.

The next step is to write
the inverse metric $g^{\mu\nu}$
as the ratio of the cofactor $C^{\mu\nu}$
over the metric determinant $g$:
\begin{equation}
\label{eq:inversemetric-ratio}
g^{\mu\nu} = \frac{C^{\mu\nu}}{g}\,.
\end{equation}
The crucial observation (already made by
Einstein and Rosen~\cite{EinsteinRosen1935})
is that this cofactor $C^{\mu\nu}$ is simply
a \emph{polynomial} in the components
of the metric $g_{\mu\nu}$.
We will give the explicit expression
for $C^{\mu\nu}$ shortly,
but let us first finish our argument.

%%\newpage%%tmp
Given the structure~\eqref{eq:inversemetric-ratio},
we see that the $g^{2}$ factor
on the left-hand side of
\eqref{eq:ext-Einstein-eq-k=2}
will precisely
cancel out the inverse factors from the
replacements~\eqref{eq:inversemetric-ratio}
for the inverse metrics in the Christoffel symbols
entering the explicit
expression~\eqref{eq:g2Rmunu-explicit}.
Indeed, in a short-hand notation,
the derivative terms $[\partial\, \Gamma]$
give $[(-g^{-2}\,\partial g)\,(\ldots)
+ g^{-1} (\partial \ldots)]$,
so that all inverse powers of $g$ are cancelled upon
multiplication by $g^{2}$ in the first two terms
on the right-hand side of \eqref{eq:g2Rmunu-explicit}.
Also, the products $[g\, \Gamma]\,[g\, \Gamma]$
in the last two terms
on the right-hand side of \eqref{eq:g2Rmunu-explicit}
give expressions of the form of
$[C\, (\ldots)]\,[C\, (\ldots)]$.
All in all, each inverse power of $g$
has disappeared and
we are left with a complicated expression
involving the
$g_{\mu\nu}$ components and their first and second
derivatives.
This completes our sketch of the proof
that $g^{2}\,R_{\mu\nu}$
does not involve the inverse metric $g^{\mu\nu}$.
(Incidentally, this simple argument also applies
to $g^{2}\,R_{\kappa\lambda\mu\nu}$, but not
to $g^{2}\,R \equiv g^{2}\,g^{\kappa\lambda}R_{\kappa\lambda}$,
which has a contribution with three inverse metrics.)

For the sake of completeness,
we now give the explicit ``recipe''
for the cofactor components $C^{\mu\nu}$
appearing in \eqref{eq:inversemetric-ratio}.
Let the matrix indices
$\overline{\mu}$ and $\overline{\nu}$
run over $\{ 0,\, 1\,, 2\,,3 \}$
and separate the indices on a matrix
with a bar instead of the usual comma.
Given the matrix  $(g)_{\,\overline{\mu}\,|\,\overline{\nu}}$
corresponding to the metric tensor $g_{\mu\nu}$,
we get the cofactor components by evaluating
the so-called Minor matrix $M$,
where the entry $(M)_{\,\overline{\mu}\,|\,\overline{\nu}}$
is obtained from the $4\times 4$ matrix
$(g)_{\,\overline{\mu}\,|\,\overline{\nu}}$
if we delete row $\overline{\mu}$
and column $\overline{\nu}$ and, then,
evaluate the determinant of the
remaining $3\times 3$ submatrix.
The final expression is obtained
if we multiply by
a sign factor $(-1)^{\overline{\mu}+\overline{\nu}}$
and transpose the Minor matrix.
[As the $(g)_{\,\overline{\mu}\,|\,\overline{\nu}}$ matrix
is symmetric, this transposition step
does not make a difference for the number obtained.]
Explicitly, we have for the cofactor components:
\begin{equation}
\label{eq:cofactor-explicit}
(C)_{\,\overline{\mu}\,|\,\overline{\nu}}
= (-1)^{\overline{\mu}+\overline{\nu}} \;
(M)_{\,\overline{\nu}\,|\,\overline{\mu}}\,,
\end{equation}
which is not to be read as a tensor equation
(left-tensor=right-tensor)
but as an expression for matrix entries.

With the components of the Minor matrix being a
polynomial in the $g_{\mu\nu}$ components
(coming from the determinants of the submatrices),
we have the result that, for degenerate
metrics with locally vanishing
metric determinant $g$,
the possible singularity of certain
inverse-metric components shows up as
a divergent factor $g^{-1}$.
Precisely these divergent factors can be
cancelled by an overall factor $g^{k}$
in the extended Einstein equation.

%%\newpage%%tmp
\subsection{RHS: Various matter components}
\label{subapp:RHS}

For the right-hand side of \eqref{eq:ext-Einstein-eq-k=2},
we need the explicit expression
of the energy-momentum tensor $T_{\mu\nu}$ and
we will start by discussing two well-known cases.

Consider, first, the case of a perfect isotropic fluid,
which has the energy-momentum tensor
(already used implicitly in the main text):
\beq\label{eq:Tmunu-perfect-fluid}
T_{\mu\nu}^\text{\,(perfect\;fluid)}=
\left(P_{M}+\rho_{M}\right)\,U_{\mu}\,U_{\nu}+P_{M}\;g_{\mu\nu}\,,
\eeq
where $U_{\mu}$ is the normalized four-velocity
of a comoving fluid element and $P_{M}$ and $\rho_{M}$
are scalars corresponding to, respectively, the pressure
and the energy density measured
in a localized inertial frame moving along with the fluid;
see, e.g., Sec.~5.4 of Ref.~\cite{Weinberg1972}.
For the metric~\eqref{eq:RWK},  %%v202
the explicit components of the normalized four-velocity
$U_{\mu}$ are as follows:
\mbox{$U_{\mu}$ $=$   %%t/\sqrt{b^{2} + t^{2}}
$\big(t \big/\big[b^{2}+t^{2}\big]^{1/2},\, 0,\, 0,\, 0\big)_{\mu}$.}

The perfect-fluid energy-momentum tensor itself does
not involve the inverse metric $g^{\mu\nu}$, but the
second term on the right-hand side of \eqref{eq:ext-Einstein-eq-k=2}
does involve the inverse metric to make for
the energy-momentum trace.
Inserting the ratio \eqref{eq:inversemetric-ratio}
for that single inverse metric eats up
one factor $g$, with the cofactor
term $C^{\mu\nu}$ remaining (i.e., a polynomial
in the $g_{\mu\nu}$ components)
and a single power of $g$ left over
from the initial factor $g^{2}$.
Hence, the right-hand side of \eqref{eq:ext-Einstein-eq-k=2} is
well-behaved for the case of a perfect fluid.
The issue of energy-momentum conservation,
also for this case, will be discussed
in App.~\ref{subapp:Energy-mom-conserv}.

Next, consider the case of a real scalar field
with Lagrange density
\begin{subequations}
\label{eq:L-scalar}
\begin{eqnarray}
\mathcal{L}^\text{(scalar)} &=&
-\frac{1}{2}\; g^{\kappa\lambda}\;
\partial_{\kappa}\phi\, \partial_{\lambda}\phi -V(\phi)\,,
\\[1mm]
V(\phi) &=&
\frac{1}{2}\;m^{2}\;\phi^{2}
+
\frac{1}{4}\;\lambda\;\phi^{4} + \ldots \,,
\end{eqnarray}
\end{subequations}
which gives the following energy-momentum tensor:
\begin{equation}
\label{eq:Tmunu-scalar}
T_{\mu\nu}^\text{\,(scalar)}=
\partial_{\mu}\phi \,\partial_{\nu}\phi
- \frac{1}{2}\; g_{\mu\nu}\;g^{\kappa\lambda}\;
\partial_{\kappa}\phi \,\partial_{\lambda}\phi
-g_{\mu\nu}\;V(\phi)\,.
\end{equation}
For the first term of the right-hand side
in \eqref{eq:ext-Einstein-eq-k=2}
we only need a single $g$ factor
to ``tame'' the manifest inverse metric
in the second term of the
energy-momentum tensor~\eqref{eq:Tmunu-scalar}.
For the second term of the
right-hand side in \eqref{eq:ext-Einstein-eq-k=2},
we then need both $g$ factors, one for
the inverse metric making the energy-momentum trace
and another for the inverse metric inside
the energy-momentum tensor~\eqref{eq:Tmunu-scalar}.
And all's well that ends well, at least for
the scalar case considered.

%%\newpage%%tmp
Of course, there are also higher-derivative
scalar theories, for which
the right-hand side of \eqref{eq:ext-Einstein-eq-k=2}
would run into trouble. One example would be a
Pais--Uhlenbeck-type
theory~\cite{PaisUhlenbeck1950}
generalized to a higher-power $n=3$ of the
d'Alembertian
\begin{equation}
\label{eq:dAlembertian-def}
\Box\equiv g^{\kappa\lambda}\;
\nabla_{\kappa}\, \nabla_{\lambda}\,,
\end{equation}
in terms of the covariant derivative $\nabla_{\kappa}$
(cf. the textbooks~\cite{Weinberg1972,HawkingEllis1973}).
The Lagrange density has been given as
Eq.~(4.29) for $n=3$ in Ref.~\cite{Gibbons-etal2019}
and reads
\begin{equation}
\label{eq:L-PU-n=3}
\mathcal{L}^{(\text{PU--scalar},\,n=3)} =
-\frac{1}{2}\;\frac{1}{M^{4}} \;
\phi \,\left(-\Box \right)^{3}\phi -V(\phi)\,,
\end{equation}
with an explicit mass scale $M$ inserted for
the correct mass dimension.
The corresponding energy-momentum tensor
is given by Eq.~(4.35) for $n=3$
in Ref.~\cite{Gibbons-etal2019}
and contains a term with three d'Alembertians:
\begin{equation}
\label{eq:Tmunu-PU-n=3}
T_{\mu\nu}^{(\text{PU--scalar},\,n=3)}=
\ldots - g_{\mu\nu}\;\Box \phi\;\Box^{2}\phi \,.
\end{equation}
This contribution thus has at least
three inverse-metric factors,
buried inside the three d'Alembertians
as defined by \eqref{eq:dAlembertian-def},
which give  at least a factor $g^{-3}$
after the replacement~\eqref{eq:inversemetric-ratio}.
(We have said here ``at least,''
because there may also be inverse
metrics from the Christoffel symbols entering
the covariant derivatives.)
Such a factor $g^{-3}$
cannot be cancelled completely by
the overall factor $g^{2}$ present in the
extended Einstein
equation~\eqref{eq:ext-Einstein-eq-k=2}.
But, most likely, the higher-derivative theory
\eqref{eq:L-PU-n=3} already runs into trouble
with unitarity~\cite{PaisUhlenbeck1950,Gibbons-etal2019}.

Perhaps it can be shown
that all consistent scalar field theories
(those that fully respect unitarity and causality)
give a satisfactory
right-hand side of \eqref{eq:ext-Einstein-eq-k=2}.
But this may not hold for higher-spin fields, such as
the electromagnetic field $A_{\mu}$
with Maxwell field strength tensor
$F_{\mu\nu} \equiv
\partial_{\mu} A_{\nu}-\partial_{\nu} A_{\mu}$
and Lagrange density
\begin{subequations}
\label{eq:L-em}
\begin{eqnarray}
\mathcal{L}^\text{(em)} &=&
-\frac{1}{4}\;
F_{\kappa\,\lambda}\,F_{\rho\sigma}\,
g^{\kappa\rho}\,g^{\lambda\sigma}\,.
\end{eqnarray}
\end{subequations}
Then, the electromagnetic energy-momentum
tensor is given by
\begin{equation}
\label{eq:Tmunu-em}
T_{\mu\nu}^\text{\,(em)}=
F_{\mu\,\lambda}\,F_{\nu\kappa}\,g^{\kappa\lambda}
-\frac{1}{4} g_{\mu\nu}\,
F_{\kappa\,\lambda}\,F_{\rho\sigma}\,
g^{\kappa\rho}\,g^{\lambda\sigma}\,,
\end{equation}
as follows from, e.g.,
Eq.~(5.3.7) in Ref.~\cite{Weinberg1972}.
The $g_{\mu\nu}$ term in \eqref{eq:Tmunu-em}
spells trouble for the second term
on the right-hand side
of \eqref{eq:ext-Einstein-eq-k=2},
as there are then three inverse metrics
giving, by the
replacements~\eqref{eq:inversemetric-ratio},
a total factor $g^{-3}$
which cannot be cancelled
completely by the explicit
$g^{2}$ factor from the right-hand side
of \eqref{eq:ext-Einstein-eq-k=2}.
Hence, we need at least $k=3$ in
\eqref{eq:ext-Einstein-eq-k=2}
for the electromagnetic case.

%%\newpage%%tmp
\subsection{Energy-momentum conservation}
\label{subapp:Energy-mom-conserv}

An interesting question has been raised~\cite{Wang2026}
as to the validity of energy-momentum conservation
at spacetime points where the metric determinant $g$
vanishes..

Assume that we have metric and matter fields for which
the $k=2$ equation \eqref{eq:ext-Einstein-eq-k=2}
holds at \emph{all} spacetime points.
Multiplication by $g$ gives the
corresponding $k=3$ equation.
Now, we can contract with
the inverse metric $g^{\mu\nu}$, interpreted
as the ratio \eqref{eq:inversemetric-ratio},
to obtain the following relation between scalars $R$
and $T \equiv g^{\kappa\lambda}\,T_{\kappa\lambda}$:
\begin{equation}
\label{eq:g3*R=g3*T}
g^{3}\,R= - 8\pi G\;g^{3}\,T\,.
\end{equation}
Then, we have from the $k=3$
equation \eqref{eq:ext-Einstein-eq-k=2}
the following alternative equation:
\begin{equation}
\label{eq:g3*Gmunu=g3*Tmunu}
g^{k}\,
\left(R_{\mu\nu}- \frac12\, g_{\mu\nu}\,R\right)
= 8\pi G\;g^{k}\,T_{\mu\nu}\,,
\qquad   k=3\,,
\end{equation}
which is everywhere well-defined
[we already noted the
need of a $g^{3}$ factor for $R$
in the parenthetical remark
``(Incidentally, $\ldots$)'' of the
pre-penultimate paragraph in
App.~\ref{subapp:LHS-Sketch-proof}].

Assuming \eqref{eq:g3*Gmunu=g3*Tmunu}
for $k=3$ to hold at all spacetime points,
we can multiply by $g$ and obtain the corresponding
equation for $k=4$.
Now, take the covariant divergence with operator
$\nabla^{\mu}$
and use the contracted Bianchi identities
\begin{equation}
\label{eq:g4*nablaMUGmunu=0}
g^{4}\;\nabla^{\mu}
\left(R_{\mu\nu}- \frac12\, g_{\mu\nu}\,R\right)
= 0\,,
\end{equation}
in order
to obtain the following conservation ``law'':
\begin{equation}
\label{eq:g4*nablaMUTmunu=0}
g^{4}\,\nabla^{\mu}T_{\mu\nu}= 0\,,
\end{equation}
based on
 the assumed validity
 of \eqref{eq:ext-Einstein-eq-k=2}
for $k=2$.

The mathematics behind \eqref{eq:g4*nablaMUTmunu=0}
is relatively straightforward,
but the physics interpretation is perhaps less clear.

Let us discuss a concrete case with the
tamed-Big-Bang metric \eqref{eq:RWK}
and a homogeneous perfect fluid having
the energy-momentum
tensor \eqref{eq:Tmunu-perfect-fluid}.
From \eqref{eq:g4*nablaMUTmunu=0},
we have
\begin{equation}
\label{eq:g4*nablaMUTmunu=0-perfect-fluid}
g_\text{RWK}^{4}\;g_\text{RWK}^{00}\,\Big[
\partial_{0}\,\rho_{M}+ 3\;\frac{\partial_{0}\,a}{a}\;
\Big(\rho_{M}+P_{M}\Big)\Big] =0\,,
\end{equation}
in terms of the derivative
$\partial_{0} \equiv \partial/\partial x^{0}
=\mathrm{d}/\mathrm{d} t$.
With $g_\text{RWK} \sim t^{2}$ and
$g_\text{RWK}^{00} \sim 1/t^{2}$ for
the metric \eqref{eq:RWK}, we then get
from \eqref{eq:g4*nablaMUTmunu=0-perfect-fluid}
\begin{equation}
\label{eq:t6*rhodot}
t^{6}\,\Bigg[
\dot{\rho}_{M}+ 3\;\frac{\dot{a}}{a}\;
\Big(\rho_{M}+P_{M}\Big)\Bigg] =0\,,
\end{equation}
where the overdot stands again for differentiation
with respect to $t$.
For a relativistic equation-of-state
parameter $w_{M}\equiv P_{M}/\rho_{M}=1/3$, this gives
\begin{equation}
\label{eq:t6*adot/a}
t^{6}\,\Bigg[\frac{\dot{a}}{a}\,\Bigg(
a\,\frac{\mathrm{d}\rho_{M}}{\mathrm{d} a}
+ 4\;\rho_{M}\Bigg)\Bigg]^{(w_{M}=1/3)} =0\,,
\end{equation}
with the solution
\begin{equation}
\label{eq:rhoM-sol}
\rho_{M}(t) \,\Big|^{(w_{M}=1/3)}
\sim a(t)^{-4}
\sim \Big( b^{2}+t^{2} \Big)^{-1}\,,
\end{equation}
as found previously
in \eqref{eq:regularized-Friedmann-asol-rhoMsol}.
The point of the exercise shown in
Eqs.~\eqref{eq:g4*nablaMUTmunu=0-perfect-fluid}--%
\eqref{eq:rhoM-sol}
is to make sure that each step
is well-defined at all spacetime
points, including those of
the hypersurface $t=0$.

For the concrete case considered,
the solution formally respects
energy-momentum conservation, but the
physical interpretation is unsatisfactory at $t=0$.
The crucial point is that
the standard elementary-flatness property
does not hold at $t=0$:
it is impossible to get a locally-flat
(Minkowskian) spacetime by
a diffeomorphism (see Sec.~II.D in
Ref.~\cite{KlinkhamerSorba2014} and
App.~D in Ref.~\cite{Klinkhamer2014-MPLAreview}).
Then, $\rho_{M}(0)$ and $P_{M}(0)$
cannot simply be interpreted as,
respectively, the ``local energy density'' and
the ``local pressure'' of the fluid, which
are only defined in a local inertial frame
comoving with the fluid,
as explained in the lines
below \eqref{eq:Tmunu-perfect-fluid}.

We conclude that, for the concrete case considered,
energy-momentum conservation formally
appears to hold, but the physical interpretation
of the involved quantities
(energy density and pressure of the fluid)
is unclear at the spacetime-defect location
($t=0$).
Essentially the same conclusion holds
for other types of matter, for example
the scalar theory \eqref{eq:L-scalar},
where the interpretation of $m$ as
a ``restmass'' would require a
locally inertial frame that, however,
cannot be reached
by a diffeomorphism at the $t=0$ defect location.
Hence, the need for an understanding
of the mechanism behind the
emergence of spacetime and matter at
the spacetime-defect location (cf. the
discussion in Sec.~6 of Ref.~\cite{Klinkhamer2025-MPLAreview}).

%%\newpage%%tmp
\subsection{Another effect: Antigravity} %%first in v2.60 as AppA4
\label{subapp:Another-effect-Antigravity}
%\setcounter{equation}{0}
%\section{Another effect: Antigravity}
%\label{app:Another-effect-Antigravity}

\subsubsection{Introduction and main result}
\label{subsubapp:Introduction-and-main-result}
%\subsection{Introduction and main result}
%\label{subapp:Introduction-and-main-result}

Assuming the validity of an extended Einstein equation,
the main new effect discussed up till now
in this paper has been the existence
of another ``side'' of the tamed Big Bang (as shown by the $t<0$
phase in Figs.~\ref{fig:a-FLRWK}
 and \ref{fig:a-FLRWK-ARROWS})
or even the existence of multiple
other sides (as shown in Fig.~\ref{fig:clover}).

Another new effect may be the phenomenon of antigravity,
which could, for example,
arise from a certain type of space-defect
embedded in a single asymptotically flat $3$-space.
A specific case is given by the $SO(3)$-Skyrmion
spacetime defect~\cite{Klinkhamer2014-prd}.
Its existence as a \underline{regular} solution
requires~\cite{KlinkhamerQueiruga2018a},
for a small enough defect length
scale $b$ and a small enough ratio of matter
energy scale over gravitational energy scale,
the appearance of a negative effective %%asymptotic
gravitational-mass leading to
``antigravity'' behavior
(other effects~\cite{KlinkhamerQueiruga2018b,%
KlinkhamerWang2018-lensing} will be mentioned
in Sec.~\ref{subsubapp:Antigravity-discussion}).
Here, ``antigravity'' is understood to correspond to
a repulsive force on a standard, positive-mass
test particle positioned far away from the defect center.
The required extended Einstein equation for the
Skyrme model is  %%, most likely,
the $k=3$ version of
\eqref{eq:ext-Einstein-eq-k=2}, as will be explained
in the last paragraph of
App.~\ref{subsubapp:Ansaetze}.

The required small enough defect length scale $b$
can be bounded parametrically as
\begin{equation}
\label{eq:b-bound-1}
b \lesssim \text{O}\left(G\,f/e\right)\,,
\end{equation}
with Newton's gravitational coupling constant $G$,
the ``pion'' energy scale $f$, and
the dimensionless coupling constant $1/e^{2}$ of the
Skyrmion term in the action
as will be given in
App.~\ref{subsubapp:Skyrme-model-for-pions}.
Recall that we use  natural units with $c=\hbar=1$.

This bound can also be written as
\begin{equation}
\label{eq:b-bound-2}
b \lesssim \left(1/e\right)\,
\left(f/E_\text{planck} \right)\,l_\text{planck}\,,
\end{equation}
with   %% \sqrt{8\pi|=5.0133
%$E_\text{Planck} \equiv 1/\sqrt{G}
%\approx 1.22\times 10^{19}\;\text{GeV}$.
$E_\text{planck} \equiv 1/\sqrt{8\pi G}
\approx 2.43\times 10^{18}\;\text{GeV}$.
and
%$l_\text{Planck} \equiv 1/E_\text{Planck}\approx
%1.62\times 10^{-35}\;\text{m}$.
$l_\text{planck} \equiv 1/E_\text{planck}\approx
8.21\times 10^{-35}\;\text{m}$.
Some illustrative numbers are as follows.
With a matter energy scale $f$ about a
factor $100$ below the gravitational energy scale
$E_\text{planck} \sim 10^{18}\;\text{GeV}$
and a rather large Skyrme quartic coupling
constant $1/e^{2} \sim 10^{6}$,
we get an antigravitating
space-defect for length scale $b$ of order
of $10 \times l_\text{planck}\sim 10^{-33}\;\text{m}$
or less.
Here, we assume that general relativity
gives an approximately reliable
description even for very small length scales
as given by \eqref{eq:b-bound-2}
with a right-hand side of order $l_\text{planck}$.

Before we give a quick derivation of this $b$ bound,
we present the necessary details of
the theory and the \textit{Ans\"{a}tze} considered,
together with a brief description of the
numerical results.

%%\newpage%%tmp
\subsubsection{Skyrme model for ``pions''}
\label{subsubapp:Skyrme-model-for-pions}
%\subsection{Skyrme model for ``pions''}
%\label{subapp:Skyrme-model-for-pions}

The theory is defined by the following action
of the metric field $g_{\mu\nu}(x)\in\mathbb{R}$ and
the scalar field $\Omega(x)\in SO(3)$:%
\bsubeqs\label{eq:action-omegamu}
\beqa\label{eq:action}
\hspace*{-13mm}
 S &\!\!=\!\!&\int_{M} d^4x\,\sqrt{-g}\,
\Bigg[
\frac{1}{16\pi G}\:R
+\frac{f^{2}}{4}\:\text{tr}\Big(\omega_\mu\,\omega^\mu\Big)
+\frac{1}{16 e^{2}}\: \text{tr}\Big(\left[\omega_\mu,\,\omega_\nu\right]
\left[\omega^\mu,\,\omega^\nu\right]\Big)\Bigg],
\\[2mm]
\label{eq:omegamu}
\hspace*{-13mm}
\omega_\mu &\!\!\equiv\!\!& \Omega^{-1}\,\partial_\mu\,\Omega\,,
\eeqa
\esubeqs
for a generic 4-dimensional spacetime manifold $M$ and
generic coordinates $x^{\mu}$, both to be
defined later. Here, $G>0$
is Newton's gravitational coupling constant,
$f>0$ the energy scale %%in the kinetic term
of the three ``pions'' contained in the
matrix field $\Omega$, and $1/e^{2}$ the
dimensionless coupling constant
of the quartic Skyrme term (a square of
commutators, with
 $[A,\,B]\equiv A\cdot B- B\cdot A$).
There are two dimensionless model parameters,
%in the theory \eqref{eq:action-omegamu},
\bsubeqs\label{eq:dimensionless-model-param}
\begin{eqnarray}\label{eq:dimensionless-eta}
\widetilde{\eta}
&\equiv& 8\pi\, G_{N}\, f^{2}
\equiv  f^{2} / E_\text{planck}^{2}
\geq 0\,,
\\[2mm]
e &\in&  (0,\,\infty)\,,
\end{eqnarray}
\esubeqs
with the gravitational energy scale
$E_\text{planck}\approx 2.44\times 10^{18}\;\text{GeV}$.

Up till now, we have written ``pions,''
where the quotation marks are to emphasize that
the particles considered may very well differ
from the usual three pions
of the strong interactions.
The reason is twofold. First, our particles
are scalars, whereas the strong-interaction
pions $\pi^{\pm,\,0}$ are known to be pseudoscalars.
Second, the energy
scale $f$ considered here may very well differ
from the strong-interaction energy
scale $f_\text{strong} = \text{O}(100\;\text{MeV})$.
From now on, we drop the quotation marks on
our pions.

%%\newpage%%tmp
\subsubsection{Ans\"{a}tze}
\label{subsubapp:Ansaetze}
%\subsection{Ans\"{a}tze}
%\label{subapp:Ansaetze}

The proper \textit{Ans\"{a}tze}
for a Skyrmion space-defect
embedded in a single Euclidean
\mbox{3-space} $E_{3}$ were obtained in
Ref.~\cite{Klinkhamer2014-prd}.
Specifically, there is, at the defect surface,
a direct antipodal identification
\big(denoted by the $\stackrel{\wedge}{=}$ symbol\big):
$b\,\widehat{x} \in S^{2} \subset E_{3}
\stackrel{\wedge}{=}  %%\Longleftrightarrow
 -b\,\widehat{x} \in S^{2} \subset E_{3}$,
 with the unit 3-vector $\widehat{x} \equiv
\vec{x}/|\vec{x}|$ from the
Cartesian coordinates of the
embedding Euclidean 3-space $E_{3}$
and the defect length scale $b>0$.
A sketch of the topology of the space-defect
will be given later in the left panel of
Fig.~\ref{fig:Skyrmion-spacetime-defect}.

The metric \textit{Ansatz} is then given by
\bsubeqs\label{eq:defect-metric-Ansatz-W-definition}
\beqa\label{eq:defect-metric-Ansatz}
\hspace*{-4mm}
ds^{2}\,\Big|_\text{defect} %%^\text{(generalized)}
&=&
- \big[\mu(W)\big]^{2}\;d t^{2}
+ \big[\sigma(W)\big]^{2}\;
  \Big(1-b^{2}/W\Big)\; d\xi^{2}
\nonumber\\[1mm]
&&
+ W\,
  \Big[ d\theta^{2} + \sin^{2}\theta\, d\phi^{2} \Big]\,,
\\[2mm]
\hspace*{-4mm}
\label{eq:W-definition}
W &\equiv& b^{2} + \xi^{2}\,,
\eeqa
\esubeqs
where $\xi$ is a quasi-radial coordinate and
the ranges of the standard spherical
coordinates $\{\theta,\, \phi\}$ are reduced
for the coordinate chart-2 considered:
\beqa\label{eq:X2Y2Z2-ranges}
\phi %%X_{2}
\in (0,\,\pi)\,,\quad
\xi %%Y_{2}
\in (-\infty,\,\infty)\,,\quad
\theta %%Z_{2}
\in (0,\,\pi)\,.
\eeqa
These chart-2 coordinates cover a region
around the $x^{2}$-Cartesian-coordinate axis and
there are two other coordinate charts
around the other two axes; see
Ref.~\cite{Klinkhamer2014-prd} for details.
The metric from
\eqref{eq:defect-metric-Ansatz-W-definition}
is degenerate, with a vanishing
determinant of the metric at the defect
surface $W=b^{2}$ with coordinate $\xi=0$.
%%and towards spatial infinity $|\xi| \to \infty$.
The $\xi=0$  defect surface corresponds, in fact,
to an $\mathbb{R}P^{2}$ manifold (especially
instructive are Figs.~3 and 4
of Ref.~\cite{Klinkhamer2014-MPLAreview},
which contains further discussion in
Secs.~2 and 3).

The boundary conditions on the metric
functions $\mu(W)$ and $\sigma(W)$
at the defect surface $W=b^{2}$\,
and at spatial infinity
($W \to \infty$ from $|\xi| \to \infty$) are:%
\bsubeqs\label{eq:sigma-mu-bc}
\beqa\label{eq:sigma-bc}
\sigma(b^{2}) &\in& (0,\,\infty)\,,
\\[2mm]
\label{eq:mu-bc}
\mu(\infty)  &=&  1\,,
\eeqa
\esubeqs
where the $\sigma$ value zero has been excluded,
in order that the field equations be well-defined
at $\xi=0$ (see Sec.~3.3.1 of Ref.~\cite{Guenther2017}).

%%\newpage%%tmp
The scalar \textit{Ansatz} is given by
%The Skyrmion-type \textit{Ansatz} for the $SO(3)$ scalar field
\bsubeqs\label{eq:hedgehog-Ansatz}
\begin{eqnarray}\label{eq:hedgehog-Ansatz-Omega}
\hspace*{-9mm}
\Omega(X) &\!=\!&
\cos\big[F(W)\big]\; \mathds{1}_{3}  %%\openone_{3} with revtex
-\sin\big[F(W)\big]\;
\widehat{x}\cdot \vec{S}
%%\nonumber\\&&
+\big(1-\cos\big[F(W)\big]\big)\;
\widehat{x} \otimes \widehat{x}\,,
\\[2mm]
\hspace*{-9mm}
S_1 &\!\equiv\!&  \left(
                \begin{array}{ccc}
                  0     & 0  &   0 \\
                  0     & 0  &   1 \\
                  \;0\; & -1 & \;0\; \\
                \end{array}
              \right)\,,
\; %%\quad
S_2 \equiv  \left(
                \begin{array}{ccc}
                  \;0\; & \;0\; & -1 \\
                    0   &   0   & 0 \\
                    1   &   0   & 0 \\
                \end{array}
              \right)\,,
\; %%\quad
S_3 \equiv  \left(
                \begin{array}{ccc}
                  0  &   1   &   0 \\
                  -1 & \;0\; & \;0\; \\
                  0  &   0   &   0 \\
                \end{array}
              \right)\,,
\\[2mm]
\hspace*{-9mm}
\label{eq:hedgehog-Ansatz-bcs}
F(b^{2}) &\!=\!& \pi\,,\quad F(\infty) = 0\,,
\end{eqnarray}
\esubeqs
with $W$ defined by \eqref{eq:W-definition}
and the unit 3-vector $\widehat{x} \equiv
\vec{x}/|\vec{x}|$ from the
Cartesian coordinates
of the
embedding Euclidean 3-space $E_{3}$
(details again in Ref.~\cite{Klinkhamer2014-prd}).
The boundary conditions \eqref{eq:hedgehog-Ansatz-bcs}
make for winding number $+1$ of the compactified map
$\overline{\Omega(X)}$ in the  3-space $E_{3}$.
Further explanations can be found in
the last paragraph of Sec.~II C in
Ref.~\cite{Klinkhamer2014-prd} and
a general discussion appears in the
text around Eq.~(9.9) of
Ref.~\cite{MantonSutcliffe2004}.

%%Spacetime
The topology of the defect 3-space is nontrivial
(a multiply-connected manifold) and the same holds
for the configuration space of the scalar matter
(total winding number equal to $+1$
from the Skyrmion configuration).

The next step is to obtain
the reduced field equations for the
metric functions $\mu(W)$ and $\sigma(W)$
and the scalar function $F(W)$.
In the original papers, this was done with
a continuous-extension
procedure~\cite{Horowitz1991,Guenther2017}
for $\xi \to 0$. But, as explained
in Sec.~\ref{subsec:ext-Einstein-eq},
it is also possible
to treat \emph{all} spacetime points \emph{equally}
if we use the extended
Einstein equation \eqref{eq:extended-Einstein-eq}.
For the case considered here, it is
the $g^{k}$-extended
Einstein equation \eqref{eq:ext-Einstein-eq-k=2}
with $k=3$.
The discussion of the last paragraph of
App.~\ref{subapp:RHS} can be taken over directly
and applied to the Skyrme-model case.
The reason is that the structure of the
Skyrme term in \eqref{eq:action-omegamu}
is analogous to that of
the Maxwell Lagrange density~\eqref{eq:L-em},
especially if we replace
the Maxwell field strength $F_{\mu\nu}$ by the
Yang--Mills field strength (which contains a
commutator).
%while the pionic kinetic term in \eqref{eq:action-omegamu},
%proportional to $f^{2}$, requires only a $g^{2}$ term.
For the record,
the $f^{2}$ kinetic term in \eqref{eq:action-omegamu}
requires only a $g^{2}$ term, as that term
is analogous to the
kinetic term of the (dimensional)
scalar fields $\phi$ in the action from
\eqref{eq:L-scalar},
which gives a $V=0$ energy-momentum tensor
\eqref{eq:Tmunu-scalar} with  at most one
inverse-metric factor.

%%\newpage%%tmp
\subsubsection{Brief description of the numerical results}
\label{subsubapp:Brief-description-numerical-results}
%\subsection{Brief description of numerical results}
%\label{subapp:Brief-description-numerical-results}

The reduced field equations
are first-order for the
metric functions $\mu(W)$ and $\sigma(W)$
and second-order for the
scalar function $F(W)$. These ODEs
are, for example, given by Eqs.~(2.8abc) in
Ref.~\cite{KlinkhamerQueiruga2018a}.
Numerical solutions with
boundary conditions \eqref{eq:sigma-mu-bc}
and \eqref{eq:hedgehog-Ansatz-bcs}
were presented in
Figs.~5-9 of that same
reference~\cite{KlinkhamerQueiruga2018a}.

For the present discussion, it is
especially instructive to write
the obtained numerical metric function $\sigma^{2}(W)$
in the form of a Schwarzschild-type factor,
\begin{equation}
\label{eq:L(W)-def}
\sigma^{2}(W) \equiv
%1/[1-L(W)/\sqrt{W}\,]
\frac{1}{1-L(W)\Big/\sqrt{W}}\,,
\end{equation}
where the asymptotic value of $L(W)$,
together with the asymptotic behavior
$\mu(W)\sigma(W) \to 1$ for $W\to\infty$,
gives the Arnowitt--Deser--Misner (ADM) mass,
\beq\label{eq:M-ADM}
M_{\text{ADM}} =
\lim_{W\rightarrow\infty}L(W)/(2G)\,.
\eeq
Actually, the numerical results
hold for dimensionless quantities
$w$ and $l(w)$, shown as lower-case letters,
if we use appropriate powers of the
combined quantity $(e\,f)$.

From the $\sigma(w)$ ODE given as Eq.~(2.8a) in
Ref.~\cite{KlinkhamerQueiruga2018a}, we derive
\beqa\label{eq:l-ODE}
\frac{\textrm{d}\,l(w)}{\textrm{d}\,w}  &=&
\widetilde{\eta}\,w^{-3/2}\;
\left(A(w)+C(w)\,
\left[w\,\frac{\textrm{d}\,F(w)}{\textrm{d}\,w}\right]^{2}
\Big/ \sigma^{2}(w) \right) \geq 0\,,
\eeqa
where the auxiliary functions $A(w)$ and $C(w)$
have been
defined in Ref.~\cite{KlinkhamerQueiruga2018a}
but are not needed
here, except for the fact that they are nonnegative.
The interpretation of the inequality \eqref{eq:l-ODE}
is that, going outwards, the mass-type variable $l(w)$
can only increase as
positive energy from the matter fields
is being added.

As remarked in the previous subsection,
we observe that the Skyrmion-space-defect
metric \eqref{eq:defect-metric-Ansatz-W-definition},
with the proper metric functions $\mu(W)$
and $\sigma(W)$ from the numerical analysis,
is a solution of the $g^{k}$-extended
Einstein equation \eqref{eq:ext-Einstein-eq-k=2}
with $k=3$.

%%\newpage%%tmp
\subsubsection{Antigravity redux}
\label{subsubapp:Antigravity-redux}
%\subsection{Antigravity redux}
%\label{subapp:Antigravity-redux}

We now give a qualitative, heuristic
explanation for the appearance
of antigravity from a Skyrmion space-defect as a regular
soliton-type solution, provided the defect length scale $b$
be small enough [length scale being set by $1/(e\,f)$,
in terms of the model parameters of the pionic matter]
and provided
the gravitational energy-scale $E_\text{planck}$
be large enough (as compared
to the energy-scale $f$ of the pion matter).
Let us assume the values of $f$ and $e$ to be fixed and positive,
and consider what happens for various values of
the gravitational coupling constant $G$.

\begin{figure*}[p]
  \centering
  \includegraphics[width=0.48\textwidth]{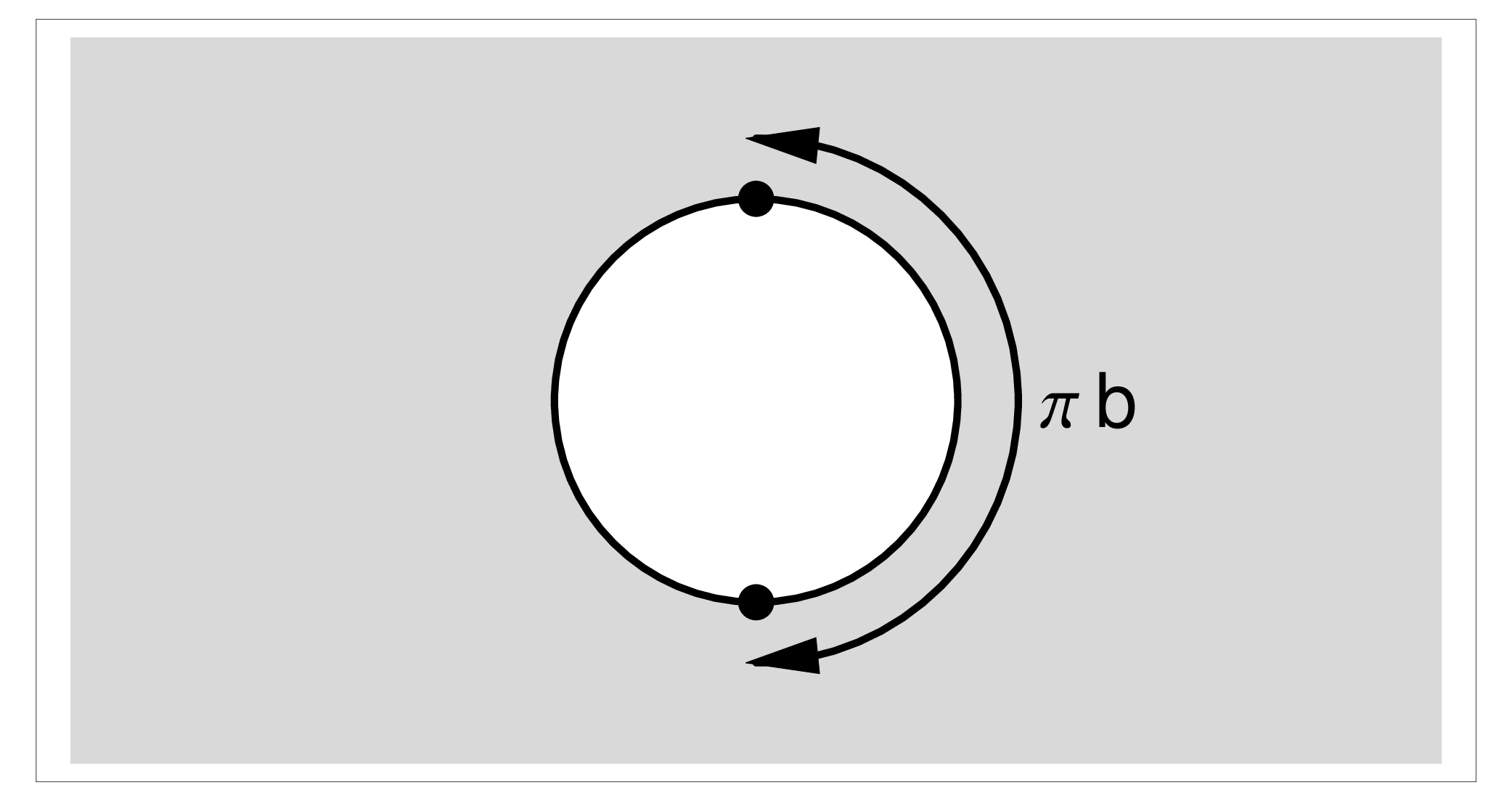}
%%{skyrmion-spacetime-defect_fig-sketch_v5.eps}
  \hfill
  \includegraphics[width=0.43\textwidth]{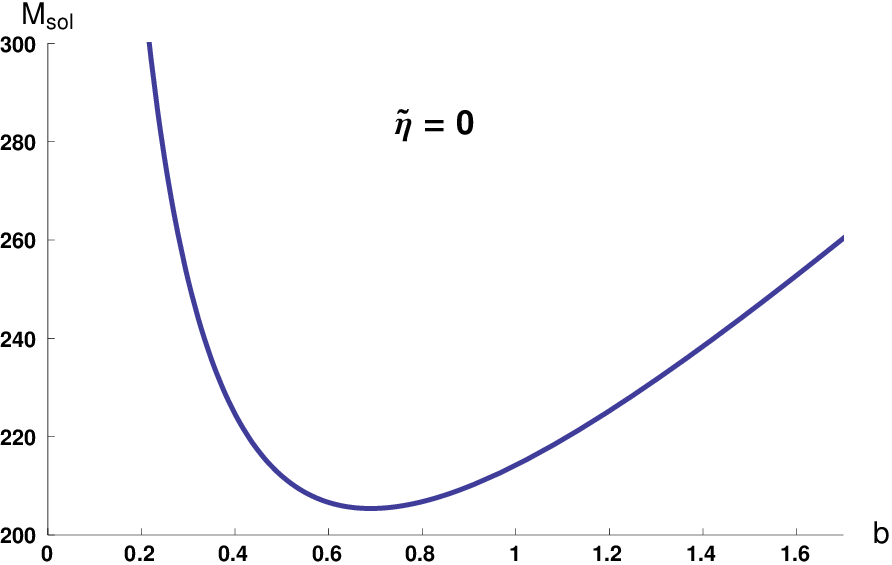}
%%{antigravity-from-a-spacetime-defect-fig1-v4.eps}
\vspace*{2pt}
\caption{Topology and mass of the
Skyrmion spacetime defect
(see the caption of
Fig.~\ref{fig:Skyrmion-spacetime-defect-num-results}
for technical details). %\newline
Left: Three-space $M_3$ obtained by surgery on
the 3-dimensional Euclidean space $E_3$:
the interior of the ball with radius $b$ is removed
and antipodal points on the boundary of the ball are
identified (as indicated by the two dots).
The ``long distance'' between antipodal points
is $\pi\, b$ in $E_3$,
while the ``short distance'' vanishes. %\newline
Right: Soliton mass $M_\text{sol}$ in units $f/e$
vs. defect  scale $b$ in units $1/(ef)$,
for the case of $G=0$ and $f,\,e > 0$,
corresponding to a dimensionless  %%model
parameter
$\widetilde{\eta}\equiv 8\pi\, G\, f^{2}=0$.
\hfill %\newline
[Image credits: left %%panel
from Fig.~1 of Ref.~\cite{Klinkhamer2014-prd} and
right %%panel
from Fig.~1 of Ref.~\cite{KlinkhamerQueiruga2018a}.]
}
\protect\label{fig:Skyrmion-spacetime-defect}
%\end{figure*}
\vspace*{0pt}
%\begin{figure*}[h]
%  \centering
\includegraphics[width=0.90\textwidth]{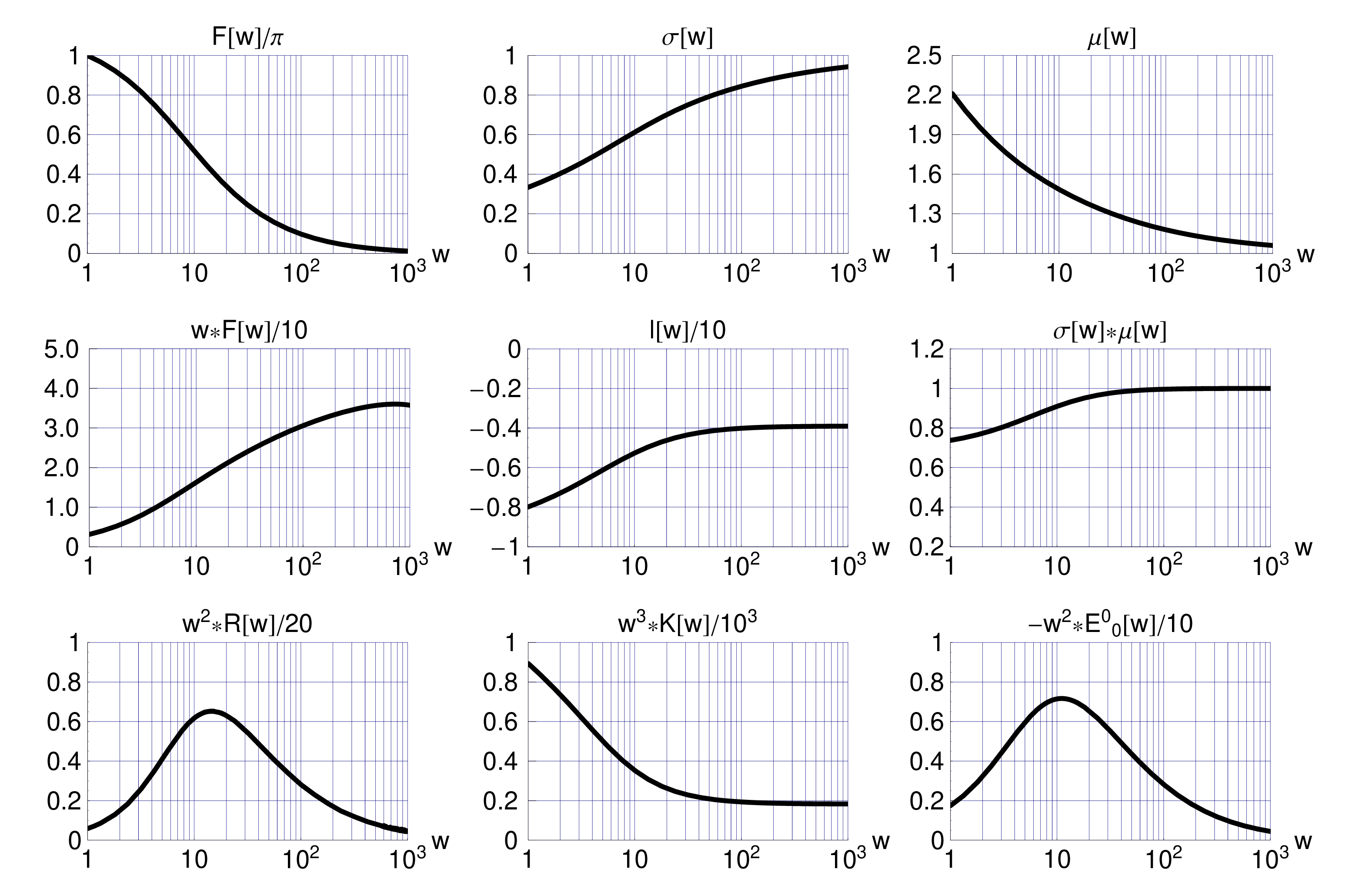}
%%{antigravity-from-a-spacetime-defect-fig5-v4.eps}
  \vspace*{1pt}
\caption{Numerical results for the
Skyrmion spacetime defect,
in terms of the  dimensionless variable
$w \equiv (e\,f)^{2}\;W$
for the coordinate $W$ from \eqref{eq:W-definition}.
For the theory \eqref{eq:action-omegamu}
and the \textit{Ans\"{a}tze}
\eqref{eq:defect-metric-Ansatz-W-definition}
and \eqref{eq:hedgehog-Ansatz},
the $k=3$ extended Einstein
equation \eqref{eq:ext-Einstein-eq-k=2}
gives the reduced field equations, which are solved
numerically~\cite{Guenther2017,%
KlinkhamerQueiruga2018a}.  %\newline
Top row:
\textit{Ansatz} functions $F(w)$, $\sigma(w)$, and $\mu(w)$
of the numerical solution
of the reduced field equations.
The parameters are $\widetilde{\eta}\equiv 8\pi\, G_{N}\, f^{2}  =1/10$ and
$y_0 \equiv ef b =1$.
The boundary conditions at the defect surface $w=(y_0)^{2}=1$ are:
$F=\pi$,
$F^\prime=-0.323978148$,
$\sigma=1/3$,
and  $\mu=2.21176$.  %\newline
Middle row: Derived functions $w\,F(w)$,
$l(w)\equiv \sqrt{w}\,\left[1-1/\sigma^{2}(w)\right]$,
and $\sigma(w)\,\mu(w)$. %\newline
Bottom row: Dimensionless Ricci curvature scalar $R(w)$,
dimensionless Kretschmann curvature  scalar $K(w)$,
and negative of the 00 component of the
dimensionless Einstein tensor $E^{\mu}_{\;\;\nu}(w)$
$\equiv$
$R^{\mu}_{\;\;\nu}(w) -(1/2)\,R(w)\,\delta^{\mu}_{\;\;\nu}$,
where the explicit expressions are given in
App.~B of Ref.~\cite{Klinkhamer2014-prd}.\hfill%%\newline
[Image credit: Fig.~5 of
Ref.~\cite{KlinkhamerQueiruga2018a}.]
}
\protect\label{fig:Skyrmion-spacetime-defect-num-results}
\end{figure*}

Without gravity ($G=0$), there is an $SO(3)$-Skyrmion over the
flat $3$-space shown in the left panel of
Fig.~\ref{fig:Skyrmion-spacetime-defect}.
The metric
\textit{Ansatz} \eqref{eq:defect-metric-Ansatz} then has
$\mu(W)=\sigma(W)=1$
and the $SO(3)$ scalar field has the Skyrmion-type
\textit{Ansatz} \eqref{eq:hedgehog-Ansatz}
 with boundary conditions for a unit-valued
winding number. The behavior of the soliton mass
$M_\text{sol}^{(G=0)}$
versus the defect length $b$ is shown in the right panel of
Fig.~\ref{fig:Skyrmion-spacetime-defect}. The soliton energy
diverges for $b\downarrow0$ due to
repulsive interaction effects from the pions, while the soliton
energy increases without bound as $b\rightarrow\infty$
due to the increasing soliton volume.

With a nonzero positive value of $G$ but still keeping
the $\mu(W)=\sigma(W)=1$
metric \eqref{eq:defect-metric-Ansatz},
we can transfer the right plot
of Fig.~\ref{fig:Skyrmion-spacetime-defect}
into a graph of distance ($b$) horizontally
versus gravitational length scale $(2 GM)$ vertically.
Multiplying the $M$ of the figure by $2G$, the crucial question
is if the parabola-like curve lies above the diagonal line or
not. If $G$ is sufficiently small, we get that the parabola-like
curve ($\propto G$) intersects with the diagonal line
($b_\text{Ordinate }=b_\text{Abscissa}$), giving
a critical value %%$b_\text{crit,\,1}$
from the smallest intersection point:
\begin{equation}
\label{eq:approx-b-crit}
2 G M_\text{sol}^{(G=0)}
\left(b_\text{crit}^\text{approx} \right)=
b_\text{crit}^\text{approx}   \,,
\end{equation}
where the suffix ``approx'' emphasizes the
approximate nature of this estimate,
the soliton mass being calculated in flat spacetime.
[We take the opportunity to correct a minor typo
in Eq.~(4.8) of Ref.~\cite{KlinkhamerQueiruga2018a},
where the $b^{2}$ dependence of
$M_{\text{sol},\,\widetilde{\eta}}$ should be a
$b$ dependence, as shown in Fig.~1 of that reference
and in the right panel
of Fig.~\ref{fig:Skyrmion-spacetime-defect} here.]

The interpretation is that, for the field configuration as given,
there will be \emph{gravitational collapse} for values
$b < b_\text{crit}$, since the defect-surface
(at $r=b$ in the corresponding Euclidean 3-space)
would lie \emph{inside} the
Schwarzschild-type radius (at $r=r_{S} \equiv 2 G M$).
There would then be no static, regular solution.

But there is a way out:
the gravitational fields of the \textit{Ansatz}
\eqref{eq:defect-metric-Ansatz} can
have a $\sigma(b^{2})$ value less than unity,
which can be interpreted as a
negative local gravitational mass according
to \eqref{eq:L(W)-def}.
Such a negative gravitational-mass contribution
to the total energy
partly cancels the large (positive)
self-interaction energy of the matter fields
with nontrivial boundary conditions $F(b^{2})=0$
and $F(\infty)=0$.

%%\newpage%%tmp
In this way, having a sufficiently negative value of the local
gravitational-mass parameter $L(b^{2})$ at the defect surface can
prevent gravitational collapse for small values of $b$.
If now, the dimensionless parameter
$\widetilde{\eta}$ from \eqref{eq:dimensionless-eta} is
very small, then the value of $L(W)$ changes very little
as $W$ increases from the value $b^{2}$ to infinity, according
to ODE~\eqref{eq:l-ODE}. In that case (small enough values of
both $b$ and $\widetilde{\eta}$),
there appears a negative ADM mass from the
definition~\eqref{eq:M-ADM}.

As said, the above discussion is qualitative
and details can be found in
Ref.~\cite{KlinkhamerQueiruga2018a}.
For example, the numerical values for
the critical defect length scale
as a function of the model parameter
$\widetilde{\eta}$ are given in
Fig.~4 of that reference.
As a further illustration of the
numerical results obtained in
Ref.~\cite{KlinkhamerQueiruga2018a}, we take
over one figure for the relatively
large value $\widetilde{\eta}=1/10$
and a defect length scale $b$ significantly
below the critical value.
Our Fig.~\ref{fig:Skyrmion-spacetime-defect-num-results}
then shows a regular solution with
metric function $\sigma(w)$ always below $1$,
so that the dimensionless mass-type parameter $l(w)$
always stays negative (cf. the middle
panel of the mid row), starting out
with a small enough value at the defect surface $w=1$
and remaining negative even though positive matter energy
is added subsequently (cf. the bottom-right panel,
with $-E^{0}_{\;\;0}$ proportional
to the matter energy density).
A case study for $\widetilde{\eta} \ll 1$
(corresponding to a matter energy scale
$f$ far below the gravitational energy scale $E_\text{planck}$)
is given in App.~A of Ref.~\cite{KlinkhamerQueiruga2018a}.

%%\newpage%%tmp
\subsubsection{Antigravity -- discussion}
\label{subsubapp:Antigravity-discussion}
%\subsection{Antigravity -- discussion}
%\label{subapp:Antigravity-discussion}

%%\vspace*{-3mm}  %%FRK workcopy

The numerical results described in
Apps.~\ref{subsubapp:Brief-description-numerical-results}
 and \ref{subsubapp:Antigravity-redux}
show that by continuously
tuning the boundary condition $\sigma(b^{2})$
from a positive value far below unity
to a  value above unity we obtain
an asymptotic gravitational mass that changes
continuously from
a negative value to a positive one.
Hence, for an appropriate value of $\sigma(b^{2})$,
we get an asymptotic
gravitational mass that vanishes exactly.

Such a defect with a vanishing asymptotic
gravitational mass has been called a ``stealth defect.''
This stealth defect still has
a localized positive energy density of the
pionic matter, which is compensated by
an effective negative gravitational energy
localized near the defect surface at $\xi=0$.
It has been studied numerically
in Ref.~\cite{KlinkhamerQueiruga2018b}.
Further phenomenological considerations
of the stealth defect were given
in Ref.~\cite{KlinkhamerWang2018-lensing},
with a special focus on gravitational-lensing effects.
We emphasize that standard gravitational lensing
is due to the curvature of spacetime that
results from ponderable matter,
whereas the lensing by the stealth defect
would be primarily due to
the nontrivial topology of the spacetime defect.
%%the matter only playing a role dynamically.

An interesting question is what the inertial mass
would be of a negative-gravitational-mass spacetime
defect or, for that matter, of a stealth defect.
Naively, we would expect a positive inertial mass,
as the matter fields contribute a positive
energy density. The violation of the
equivalence principle is perhaps to be expected
for the degenerate metric considered;
cf. the discussion in Sec.~5.1 of
Ref.~\cite{Guenther2017}.
Anyway, a definite answer to the
inertial-mass question would appear
to require, for example,
the construction of solutions of
the $k=3$ extended Einstein
equation \eqref{eq:ext-Einstein-eq-k=2}
corresponding to multiple Skyrmion
spacetime defects, initially widely separated
and with relative motions (all going outwards
or some going inwards for a
scattering subprocess). Obviously,
such a numerical calculation would
require a substantial effort;
cf. the discussion of Sec.~9.8
in Ref.~\cite{MantonSutcliffe2004}
for a fixed Minkowski spacetime.
% remains an outstanding task.

It is a straightforward exercise to obtain
a defect-wormhole version of the
Skyrmion \mbox{3-space} defect.
The proper \textit{Ans\"{a}tze}
for the defect-wormhole version
are obtained
by extending the Skyrmion-spacetime-defect
structure~\cite{Klinkhamer2014-prd} embedded
in a \emph{single} Euclidean 3-space $E_{3}$
to the
defect-wormhole structure~\cite{Klinkhamer2023a}
that connects \emph{two} asymptotically
flat 3-spaces $M_{3}^{(\pm)}$ (where the $\pm$ suffix
distinguishes the orientation, left-handed vs. right-handed).
Specifically, the direct antipodal
identification at the defect surface of the
Skyrmion spacetime defect
\big($b\,\widehat{x} \in S^{2} \subset E_{3}
\stackrel{\wedge}{=}  %%\Longleftrightarrow
 -b\,\widehat{x} \in S^{2} \subset E_{3}$\big)
 is changed into an
quasi-antipodal identification
at the defect surface of the wormhole
\big($b\,\widehat{x} \in S^{2}_{+} \subset M_{3}^{(+)}
\stackrel{\wedge}{=} %%\Longleftrightarrow
-b\,\widehat{x} \in S^{2}_{-} \subset M_{3}^{(-)}$\big).

The resulting Skyrmion-anti-Skyrmion
defect wormhole can give rise to
antigravity for small enough wormhole scale $b$
and small enough energy-scales ratio $f/E_\text{planck}$,
as discussed
in App.~\ref{subsubapp:Introduction-and-main-result}
for the Skyrmion defect.
But the topology of the wormhole
3-space is essentially trivial
(a simply connected manifold) and the same holds
for the configuration space of the
scalar matter
\big(total winding number equal to $+1-1=0$, with
a Skyrmion in the space $M_{3}^{(+)}$
and an anti-Skyrmion in the space $M_{3}^{(-)}$\big).
This makes the global stability of the defect-wormhole
configuration doubtful.

%%\newpage%%tmp
Returning to the Skyrmion spacetime defect
(but keeping in mind the possibility of a related
wormhole), we then have
an explicit mechanism for the appearance of antigravity.
This mechanism essentially
results from the nontrivial gravitational fields
at the defect surface for an appropriate degenerate metric
and assumes the validity of the $k=3$ extended Einstein
equation \eqref{eq:ext-Einstein-eq-k=2} for the
Skyrme model considered.
With this theoretical understanding in hand,
the obvious question is
if negative-gravitational-mass
objects exist in the actual Universe.
%A recent paper~\cite{NojiriOdintsov2026}
%reviews some options, but leaves the
%answer entirely open.
At this moment, we know of no suitable
astronomical candidate object and we refer
to, e.g., Ref.~\cite{Chardin2022} for
further discussion of some experimental issues.

%%\newpage%%tmp

\end{document}